\documentclass[twocolumn]{aa}

\usepackage{graphicx}
\usepackage{url}
\usepackage{txfonts}
\usepackage{natbib}

\bibpunct{(}{)}{;}{a}{}{,}
\usepackage{hyperref}

\usepackage{color}

\def\XMM{XMM-\textit{Newton}\ }

\begin{document}
    \title{\textit{F}-VIPGI: a new adapted version of VIPGI for FORS2 spectroscopy}
    \subtitle{Application to a sample of 16 X-ray selected galaxy clusters at $0.6 \leq z \leq 1.2$}

   \author{A. Nastasi
          \inst{\ref{MPE}}
          \and
	  M. Scodeggio
	  \inst{\ref{IASF}}
	  \and
 	  R. Fassbender
           \inst{\ref{MPE}}
           \and
          H. B\"ohringer
          \inst{\ref{MPE}}
          \and
	  D. Pierini
 	  \inst{\ref{FLA}}
	  \and
	  M. Verdugo
	  \inst{\ref{AIWien}}
	  B. Garilli
	  \inst{\ref{IASF}}
	  \and
	  P. Franzetti 
	  \inst{\ref{IASF}}
          }

  \institute{Max-Planck-Institut f\"ur extraterrestrische Physik (MPE),
              Giessenbachstrasse~1, 85748 Garching, Germany \\
              \email{alessandro.nastasi@mpe.mpg.de} \label{MPE}
            \and
	  IASF-INAF - Via Bassini 15, I-20133, Milano, Italy \label{IASF}
	  \and
	  Institut f\"ur Astrophysik Universit\"at Wien, T\"urkenschanzstr. 17, 1180 Vienna, Austria \label{AIWien}
	\and
	Freelance scientist \label{FLA}
             }

   \date{Received ...; accepted ...}

  \abstract
    {}
   {The goal of this paper is twofold. Firstly, we present \textit{F}-VIPGI, a new version of the VIMOS Interactive Pipeline and Graphical Interface (VIPGI) adapted to handle FORS2 spectroscopic data taken with the standard instrument configuration. Secondly, we investigate the spectro-photometric properties of a sample of galaxies residing in distant X-ray selected galaxy clusters, the optical spectra of which were reduced with this new pipeline.}
   {We provide basic technical information about the innovations of the new software and refer the reader to the original VIPGI paper for a detailed description of the core functions and performances. As a demonstration of the capabilities of the new pipeline, we then show results obtained for 16 distant ($0.65 \leq z \leq 1.25$) X-ray luminous galaxy clusters selected within the \XMM Distant Cluster Project. We performed a spectral indices analysis of the extracted optical spectra of their members, based on which we created a library of composite high signal-to-noise ratio spectra. We then compared the average spectra of the passive galaxies of our sample with those computed for the same class of objects that reside in the field at similar high redshift and in groups in the local Universe. Finally, We computed the ``photometric'' properties of our templates and compared them with those of the Coma Cluster galaxies, which we took as representative of the local cluster population.}
   {We demonstrate the capabilities of \textit{F}-VIPGI, whose strength is an increased efficiency and a simultaneous shortening of FORS2 spectroscopic data reduction time by a factor of $\sim$10 w.r.t. the standard IRAF procedures. We then discuss the quality of the final stacked optical spectra and provide them in electronic form\thanks{The library of spectra is available at the CDS via anonymous ftp to \texttt{cdsarc.u-strasbg.fr 130.79.128.5} or via \href{http://cdsweb.u-strasbg.fr/cgi-bin/qcat?J/A+A/}{http://cdsweb.u-strasbg.fr/cgi-bin/qcat?J/A+A/} and at the link provided in Appendix\,\ref{par:Appendix_A} (Sect.\,\ref{par:Appendix_A_getFVIPGI}).} as high-quality spectral templates, representative of passive and star-forming galaxies residing in distant galaxy clusters. By comparing the spectro-photometric properties of our templates with the local and distant galaxy population residing in different environments, we find that passive galaxies in clusters appear to be well evolved already at $z \sim$ 0.8 and even more so than the field galaxies at similar redshift. Even though these findings would point toward a significant acceleration of galaxy evolution in densest environments, we cannot exclude the importance of the \textit{mass} as the main evolutionary driving element either. The latter effect may indeed be justified by the similarity of our composite passive spectrum with the luminous red galaxies template at intermediate redshift.}
   {}


   \authorrunning{A. Nastasi et al.}

   \maketitle


\section{Introduction}
\label{par:intro}
Galaxy clusters are the signature of the primordial density fluctuations that have grown via hierarchical accretion since the epoch of recombination. Because their abundance at different epochs is extremely sensitive to the matter content and acceleration of the universe, clusters are sensitive probes for testing different cosmological models. In addition, they are cosmic laboratories in which complex processes that shape galaxy evolution can be studied in great detail.
Nowadays many efforts are invested into the observational challenge of providing sizable samples of galaxy clusters at high redshift ($z >$\,0.8) to trace the evolution of the cluster population and its matter components back to the first half of the universe lifetime, corresponding to lookback times of 7 $-$ 10 Gyr. Many surveys have been designed to efficiently detect distant clusters by means of their red galaxy population \citep[e.g., \textit{SpARCS},][]{Wilson2006}, their Sunyaev$-$Zel'dovich (SZ) effect signature (e.g., \textit{SPT}, \cite{Williamson2011}; \textit{ACT}, \cite{Menanteau2010}) or the diffuse X-ray emission originating from the hot intracluster medium. The last approach, in particular, has proved very powerful to the above aim as shown e.g. by the \textit{XMM}-Newton Distant Cluster Project \cite[\textit{XDCP},][]{HxB2005, Fassbender2011b}. This is a serendipitous X-ray survey specifically designed for finding and studying distant X-ray luminous galaxy clusters at $z \geq$ 0.8 and it has compiled the largest sample of such systems to date. For a comprehensive overview of the survey and an extensive discussion on its strategy and results we refer the reader to \cite{Fassbender2011b}.

Irrespective of the initial approach used to detect distant clusters, the final mandatory step of the confirmation process is the redshift assessment of the system by means of spectroscopic observations of its galaxy population. To maximize the number of galaxies at $z >$ 0.8 whose redshift can be successfully measured, one has to design spectroscopic campaigns to observe the spectral features with the highest signal-to-noise ratio (S/N). Because galaxy populations in clusters are dominated by red, passive galaxies, the most suitable feature to this aim is the continuum break at 4000\AA{} (D4000) and the associated absorption calcium lines (CaH/K). For $z \sim$ 0.8 $-$ 1 this spectral region is redshifted to $\lambda \approx$ 7200 $-$ 8000\AA{}, a wavelength window readily accessible for many ground-based telescopes. In particular, one of the currently most efficient spectrographs able to cover this wavelength range and to simultaneously guarantee a very high and stable efficiency of its CCD up to $\lambda \sim$ 11000\AA{} is the FOcal Reducer and low dispersion Spectrograph (\textit{FORS2}, Sect.\,\ref{par:fors2}) mounted on the UT1 of VLT. Because of its performances at longer wavelengths, this instrument is widely used for imaging and spectroscopy of distant cluster galaxies and is hence the instrument of choice for the spectroscopic follow-up of the distant candidate systems in XDCP.

At the time of writing XDCP provides the largest sample of X-ray selected distant galaxy clusters, with 30 confirmed clusters at $z >$ 0.9 and a final aim of more than 50 clusters at $z >$ 0.8 (30 at $z >$ 1) to allow statistically meaningful evolution studies of the cluster population in at least three mass- and redshift bins. One of the main disadvantages connected to these expectations is the significant work related to reducing the large amount of data produced by the spectroscopic follow-up campaigns. Each XDCP target is observed with the FORS2 multi-object spectroscopy (MOS) mode, enabling an average of 50 slits per mask. According to the survey expectations quoted above, the spectroscopic campaigns should finally yield several thousand slits, resulting in a similar amount of spectra to be reduced and extracted (see Fig.\,\ref{fig:wavCalRMS}, top). This process is very time-consuming if carried out with the traditional IRAF packages, and the pipelines provided by ESO for automating the procedures would not provide the necessary accuracy for an efficient extraction of the spectra of such distant galaxies, which are mostly faint (I $>$ 20)\footnote{Apparent Vega I-band magnitude of a $z >$ 0.5 L$^{\star}$ passively evolving galaxy with formation redshift $z_{form}$=5 and solar metallicity.}.
These reasons motivated us to develop a new dedicated pipeline for a quick and efficient reduction of spectroscopic FORS2 data. Because of the many similarities between FORS2 and VIMOS data, we chose to build the new pipeline on \textit{VIPGI} \citep{Scodeggio2005} and named it \textit{F}-VIPGI.

This paper is structured as follows: in Sect.\,\ref{par:vipgi} the main characteristics of VIPGI are briefly described to introduce the innovations of \textit{F}-VIPGI, which are extensively discussed in Sect.\,\ref{par:fvipgi}. We then show an application of the new pipeline to a sample of distant clusters (Sect.\,\ref{par:application}) that results in a new libray of spectroscopic templates, while technical details and properties are discussed in Sect.\,\ref{par:library} and a conclusive summary is given in Sect.\,\ref{par:discussion}.
Throughout this manuscript we assume a concordance $\Lambda$CDM cosmology, with $H_0 =$ 70 km s$^{-1}$ Mpc$^{-1}$, $\Omega_\Lambda =$ 0.7, $\Omega_m =$ 0.3 and $w=-1$.

\section{The VIMOS Interactive Pipeline and Graphical Interface (VIPGI)}
\label{par:vipgi}
VIPGI is a semi-automatic data reduction pipeline released in 2005 \citep{Scodeggio2005} written to process and archive the data obtained with the \textit{VIsible MultiObject Spectrograph} (VIMOS) mounted at the Melipal Unit Telescope (UT3) of the VLT in a quick and efficient fashion. It is able to handle the data taken with all the three available VIMOS modes: MOS, imaging, and integral field unit (IFU) spectroscopy.
The creation of such a new, VIMOS-dedicated, reduction pipeline was motivated by the need to process the huge amount of spectroscopic data expected from surveys such as the VIMOS-VLT Deep Survey \citep[VVDS,][]{LeFevre2004} or zCOSMOS \citep{Lilly2007}, each of which produced a total of $\sim$50,000 spectra of galaxies in a very wide redshift range ($0 \leq z \leq 5$).

The core of VIPGI is a library of C-written routines, the VIMOS Data Reduction Software \citep[DRS,][]{Scodeggio2001}, currently used by ESO for the VIMOS online reduction. All these fundamental functions can interact with each other and with the user thanks to the adopted standard Python Tkinter graphical interface. 
This choice allows one to obtain a pipeline where the power and efficiency of the C code computation and the possibility of a continuous quality check by the user on the intermediate reduction results are both present. Namely, the user can constantly monitor the quality of the onging reduction step by step with VIPGI, repeating an intermediate reduction step if necessary without restarting the entire procedure.

VIPGI was also designed to optimize the storage of a large amount of reduced data, with a clear and easily understandable filing strategy. More technical details on VIPGI are provided in the original paper of \cite{Scodeggio2005}.

\section{The FORS2-VIMOS Interactive Pipeline and Graphical Interface (F-VIPGI)}
\label{par:fvipgi}
Although VIPGI was specifically designed for the VIMOS instrument, its capabilities are general enough to make it potentially usable with any other MOS spectrograph. This motivated us to consider VIPGI as a potentially useful tool for XDCP, whose cluster candidates have been (or are planned to be) spectroscopically followed-up to safely confirm their nature of gravitationally bound, distant systems. The instrument used for this purpose is the \textit{FOcal Reducer and low dispersion Spectrograph} (FORS2) mounted on VLT.
The XDCP spectroscopic campaigns have targeted $\sim$70 cluster candidates since 2005, producing a total of $\sim$3,500 single spectra to be extracted and reduced. With the same original spirit of VIPGI we therefore adapted the power and efficiency of VIPGI to FORS2 data as well. The result is \textit{F}-VIPGI, a new pipeline that allows us to shorten the time for reducing FORS2 spectroscopic data by a factor of $\sim$\textit{10} w.r.t. the standard IRAF procedures.

\subsection{The FORS2 instrument}
\label{par:fors2}
FORS2 is the visual and near-UV FOcal Reducer and low-dispersion Spectrograph mounted on the UT1 unit (Antu) of the Very Large Telescope (VLT) \citep{Appenzeller1998}. The instrument covers the wavelength range from 330 nm to 1100 nm with an image scale of 0.25\arcsec/pixel (or 0.125\arcsec/pixel if the high-resolution collimator is used) in the standard readout mode (2x2 binning) and a field of view (FoV) of 6.83\arcmin $\times$6.83\arcmin. The detector consists of two MIT/LL CCID-20 chips, with 4096$\times$2048 15$\mu$m pixels, each characterized by an excellent sensitivity toward the red part of the spectrum (up to $\lambda \sim$ 11000\AA{}) and the almost total absence of fringing pattern contamination. FORS2 can be used in many modes, including multi-object spectroscopy with exchangable masks, long-slit spectroscopy, imaging, spectro-polarimetry and high-time resolution imaging and spectroscopy. 

\textit{F}-VIPGI was designed to work with all FORS2 data taken with the standard spectroscopic instrument configuration and straight slits, defined as those slits where the pixel-to-wavelength relation is ``constant'' along the slit length, and therefore the sky lines are perfectly aligned along the CCD rows (or columns, depending on the original orientation of the data). A summary of the properties of the FORS2 standard grisms equipment is provided in Table\,\ref{Tab:fors2GRISM}.
\begin{table*}[t] 
\caption{Properties of the grism forming the FORS2 standard instrument configuration for which \textit{F}-VIPGI is usable. In parenthesis are the wavelength ranges actually used by \textit{F}-VIPGI to provide the best-quality results. The listed values of resolution $\lambda/\Delta\lambda$ are computed at the central wavelength and for a 1\arcsec\ slit.}
\centering
\label{Tab:fors2GRISM}     
\begin{tabular}{l c c c c c c}
    \vspace{-0.3cm} \\
    \hline\hline \vspace{-0.3cm}            \\              
    Grism name	& Central wavelength	& Wavelength range	& \multicolumn{2}{c}{Dispersion}& 	Resolution		& Order separation \\
		    &	[nm]		&	[nm]		&  [\AA{}/mm]	&[\AA{}/pixel]	& $ \lambda / \Delta\lambda $	&	filter	 \\
    \vspace{-0.3cm}            \\
    \hline \vspace{-0.3cm}            \\
    \vspace{-0.3cm} \\
    GRIS\_600B+22	&	465		&      330 - 621	&	50	&	1.50	&	780			&	none		\\
    GRIS\_300V+10	&	590		& 330 (350) - 660 (925)	&	112	&	3.36	&	440			&	none		\\
    GRIS\_300V+10	&	590		&445 (450) - 865 (850)  &	112	&	3.36	&	440			&	GG435+81	\\
    GRIS\_300I+11	&	860		&\,\,\,\,\,\,580 (600) - 1100 (1050)&	108	&	3.24	&	660			&	none		\\
    GRIS\_300I+11	&	860		&\hspace{1.07cm}600 - 1100 (1050)&	108	&	3.24	&	660			&	OG590+32	\\
    GRIS\_150I+27	&	720		& 330 (370) - 650 (980)	&	230	&	6.90	&	260			&	none		\\
    GRIS\_150I+27	&	720		&445 (430) - 880 (990) &	230	&	6.90	&	260			&	GG435+81	\\
    GRIS\_150I+27	&	720		&\,\,\,\,\,\,600 (590) - 1100 (1050)&	230	&	6.90	&	260			&	OG590+32	\\
    \vspace{-0.3cm}            \\
\hline \hline                                   
\end{tabular}
\end{table*}

\subsection{Conversion of FORS2 files into the VIMOS format}
\label{par:fors2conversion}
The functionalities of \textit{F}-VIPGI were enabled not by creating new routines specifically coded for the FORS2 data format but instead by \textit{manipulating} the FORS2 data themselves to convert them into the VIMOS format. Despite the differences between FORS2 and VIMOS CCDs architectures (with four squared sensors for the former and two rectangular ones for the latter), our idea was to make the software able to identify chips 1 and 2 (hereafter Q1 and Q2, respectively) of FORS2 as the first two individual quadrants of VIMOS. In this way one can use the standard VIPGI recipes with the ``adapted'' FORS2 fits files without affecting the quality and reliability of the data reduction results.
The parts of FORS2 data that are manipulated by \textit{F}-VIPGI are the \textit{image} and the \textit{header}:
\begin{itemize}
\item[$\bullet$] \textit{image}: the original FORS2 matrix of 4096 $\times$ 2048 pixels is \textit{transposed} to obtain the same vertical orientation of the VIMOS frames. The wavelength dispersion orientation from lower (blue) to upper (red) side of the image is recovered in the same way. We also stress that this process does not affect the quality of the data because only the positions of the pixels are rearranged, whereas their information (counts) are left untouched.
\item[$\bullet$] \textit{header}: the header structure of FORS2 files differs significantly from those of VIMOS. In particular, while in each VIMOS file the header contains only the information on the slits of the chip to which the file belongs, in FORS2 each file has the information of all the slits of the mask on its header that are associated with Q1 and Q2. During the conversion process \textit{F}-VIPGI therefore changes these settings and removes all the information about the slits that do not belong to the chip of the frame itself. Only the six reference slits, common to both chips, are preserved along this process. We highlight that these changes are made for all FORS2 frames except the BIAS, as they do not contain any slit information in their headers.
\end{itemize}

A specific and slightly different treatment is reserved to standard-star frames (STD), as discussed in Sect.\,\ref{par:SPHcalib}.

\subsection{\textit{F}-VIPGI calibration files}
\label{par:fors2Calibration}
As for VIPGI, the user has to initially provide specific calibration files also for \textit{F}-VIPGI. These are then used by the pipeline to locate the position and length of the slits on the CCD, to compute the coefficient of the interpolating polynomyal for the wavelength dispersion, to correct for bad CCD pixels, and to finally extract the single 1-D spectra.
The calibration files are the following:
\begin{itemize}
\item[$\bullet$] \textit{grism table}: it contains all the main spectroscopic information of the grism, i.e. central wavelength, wavelength range, and resolution. The emission sky lines usable for an additional refinement of the wavelength calibration (see Sect.\,\ref{par:wavCalibration}) are defined here, too. It is defined for each grism configuration.
\item[$\bullet$] \textit{CCD table}: it maps all bad pixels/columns of the CCD that have to be corrected for by the pipeline. Two different files are needed for Q1 and Q2, but these remain the same for any FORS2 observation.
\item[$\bullet$] \textit{line catalog}: it contains the list of all arc lines used for the wavelength calibration (Sect.\,\ref{par:wavCalibration}). For FORS2 only three files are needed, depending on the kind of lamp used (HeAr, HgCdHeAr or HgCdHeNeAr).
\item[$\bullet$] \textit{grism PAF file}: it is a parameter file in the standard ESO data interface control terminology. Basically, it is just a text file where information can be collected in a format that is very similar to the FITS keywords format and is then used to update FITS keywords in the header of FITS files. Specifically, this file contains all information related to the wavelength dispersion, optical distortion, and curvature of each CCD and each grism.
\item[$\bullet$] \textit{SPHOT table}: it is a fits table containing magnitude vs. wavelength of the standard star that is used for the spectrophotometric calibration of the data. It has to be named exactly as the target name reported in the header keyword \textquotedblleft\texttt{ESO OBS TARG NAME}\textquotedblright.
\end{itemize}

All the above calibration files except for the SPHOT table are provided together with the installation package and should not be modified by the user.

\subsection{Wavelength calibration}
\label{par:wavCalibration}
\begin{figure}[t]
 \centering
\sidecaption
\includegraphics[height=7.8cm, clip=true]{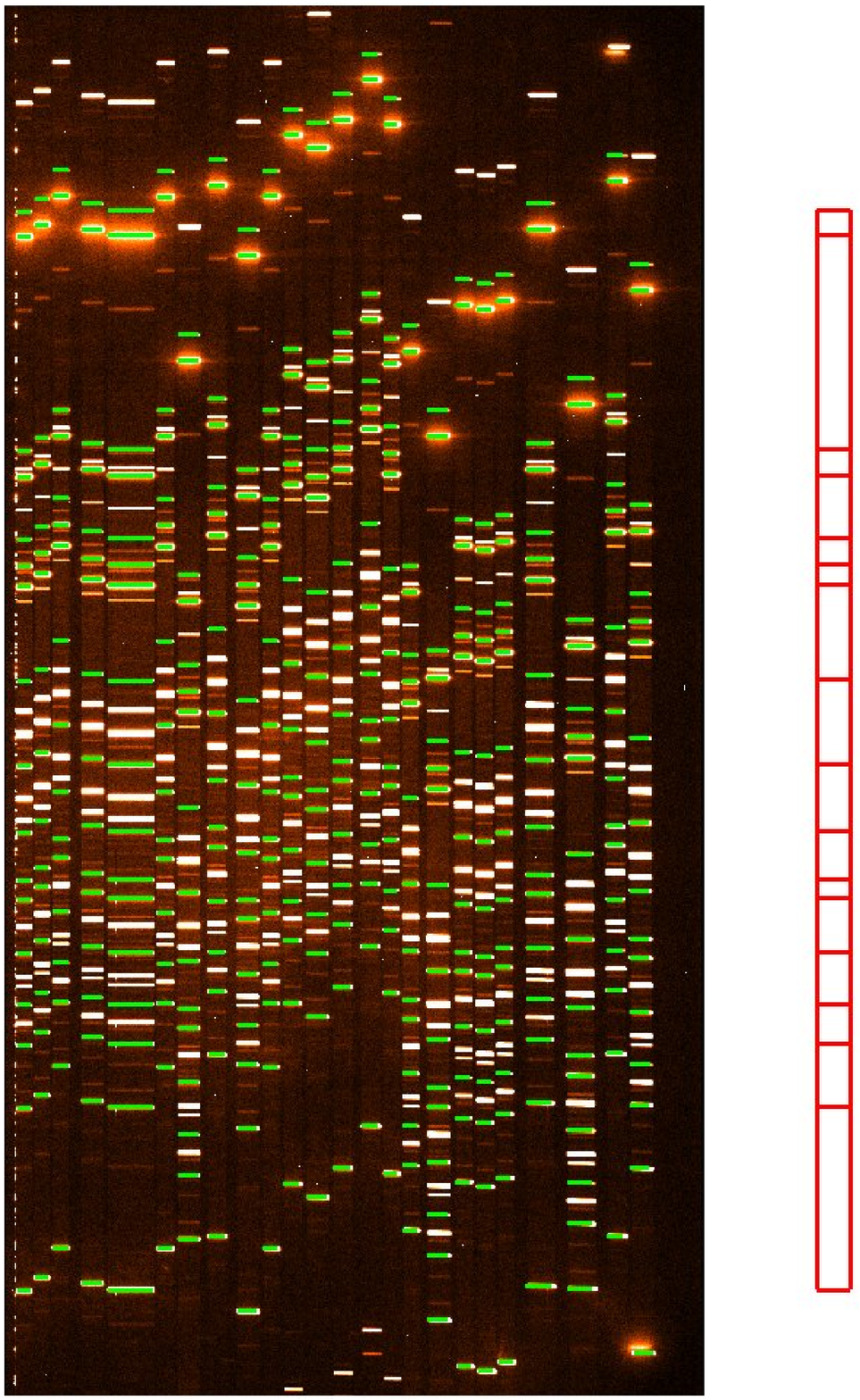} 
\caption{An example of a HeAr arc line catalog displayed on a raw lamp frame (in background) within \textit{F}-VIPGI. The green regions mark the expected positions of the arc lines according to the information extracted from the fits header while the red ones on the right represent the line catalog used for the calibration. The red part of the spectrum is in the upper part of the image.}
\label{fig:completeAdjustment}
\end{figure}

One of the most powerful aspects of VIPGI, and hence of \textit{F}-VIPGI, is that the user can perform sensitive calibration steps via visual tools that facilitate executing and constantly checking all the procedures.
An example of this interactivity within the reduction process is shown in Fig.\,\ref{fig:completeAdjustment}. Here the pipeline shows the expected positions of the arc lines of each slit (green regions overlaid on the raw lamp frame) together with the line catalog used for the calibration (red regions on the right side). The user can iteratively refine the wavelength calibration across the entire chip by shifting each green line until the region line patterns perfectly match the underlying lamp frame. The positions of the line centroids are then computed automatically by the pipeline by applying a 2.5 sigma iterative rejection to remove lines that are deviate too far from the fit.

\begin{figure}[ht]
 \centering
\includegraphics[height=7.5cm, clip=true]{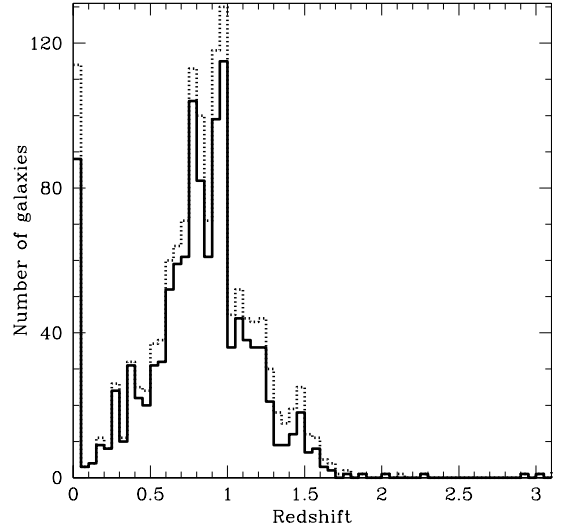} 
\includegraphics[height=7.5cm, clip=true]{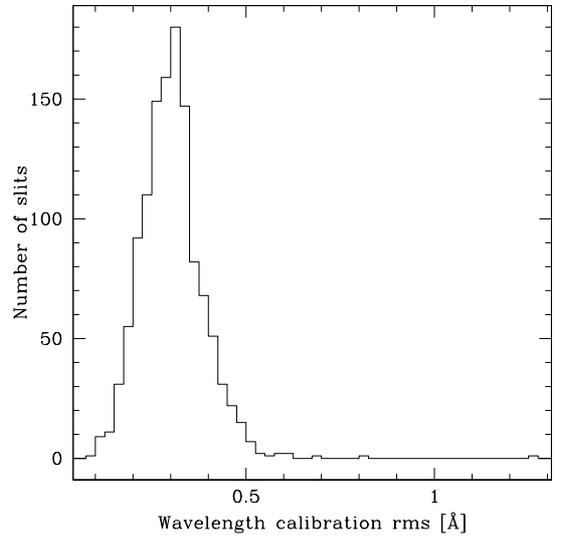} 
\caption{\textit{Top:} Redshift distribution of 1543 FORS2 spectra produced by XDCP and extracted with \textit{F}-VIPGI. The solid line marks the secure assigned redshifts ($\geq$75\% confidence on the assigned value) while the dotted line refers to unsafe ones (confidence $\leq$50\%). The secure assessments represent $\sim$72\% of the entire sample at $0 < z \leq 3$ and increases to $\sim$78\% including also stars. \textit{Bottom:} Distribution of rms wavelength calibrations computed for the corresponding 1230 reduced slits observed with the grism 300I+11. The median value is $< \mbox{rms} >_{\lambda} =$ (0.30 $\pm$ 0.05)\AA{}.}
\label{fig:wavCalRMS}
\end{figure}

This method produces a very accurate calibration with a median uncertainty on the wavelength calibration that roughly corresponds to one fifth of a pixel. As an example, for grism 300I+11, whose spectra have a  linear dispersion of 3.24\AA{}/pixel, the typical uncertainty on the wavelength calibration has an $< \mbox{rms} >_{\lambda} =$ (0.30 $\pm$ 0.05)\AA{}, as shown in the bottom panel of Fig.\,\ref{fig:wavCalRMS}.
Another refinement of the calibration can also be obtained by using the position of some bright relatively isolated sky emission lines whose wavelengths are known and defined in the \textit{grism table} (see Sect.\,\ref{par:fors2Calibration}). For data taken with grism 300I+11 we used eight sky lines between $\lambda$ = 6300\AA{} and $\lambda$ = 10120\AA{}, which produced a final refined wavelength calibration with a typical $< \mbox{rms} >_{\lambda}\ < 1$\AA{}. This quantity can be converted into a corresponding redshift uncertainty of $\delta z_{\lambda} \sim 1 - 2 \cdot 10^{-4}$ by referring to $\lambda_c$ = 8600\AA{}.
After the calibration process, all data are preliminarily reduced in the standard way by subtracting the bias frames, correcting for the CCD bad pixels and, finally, applying the flat-fielding correction.

\subsection{Sky line subtraction and atmospheric absorption corrections}
\label{par:telluric}
\begin{figure*}[t]
\centering
\includegraphics[width=18cm, clip=true]{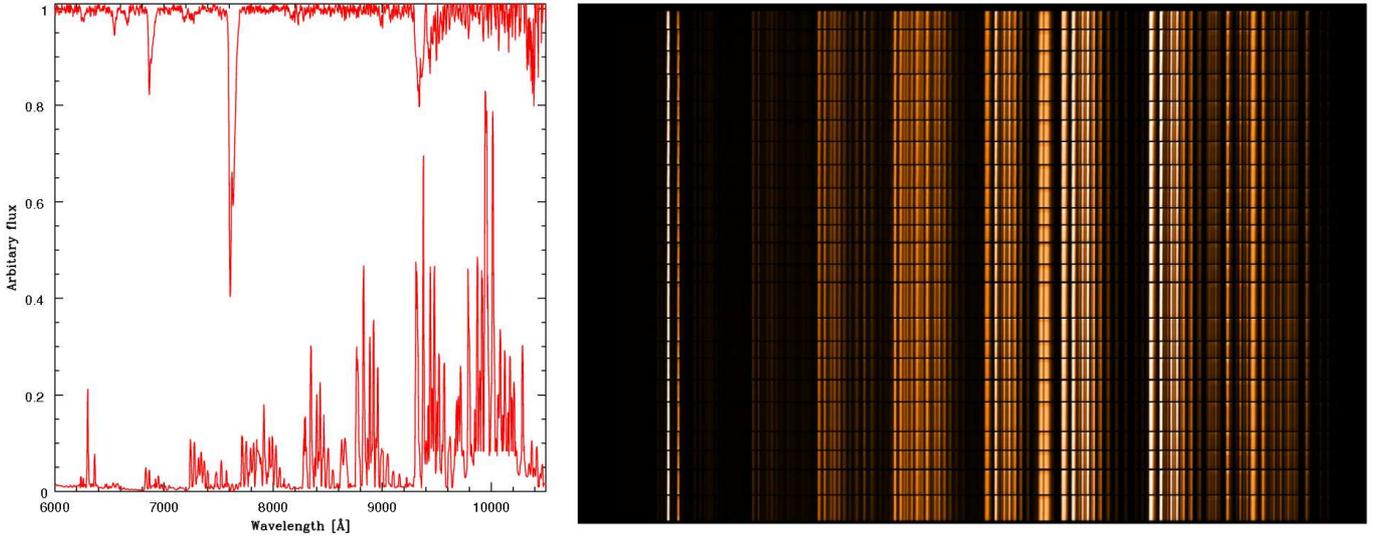} 
\caption{\textit{Left:} Spectral absorption (top side) and emission (bottom side) features introduced by the night sky in the $\lambda >$ 6000\AA{} spectral region. \textit{Right:} 2D emission spectra of the sky extracted for all slits of a FORS2 mask. The vertical alignment of the sky lines is indicative of a very good wavelength calibration over the entire mask.}
\label{fig:SkyTell}
\end{figure*}

\begin{figure*}[t]
\centering
\includegraphics[width=18cm, clip=true]{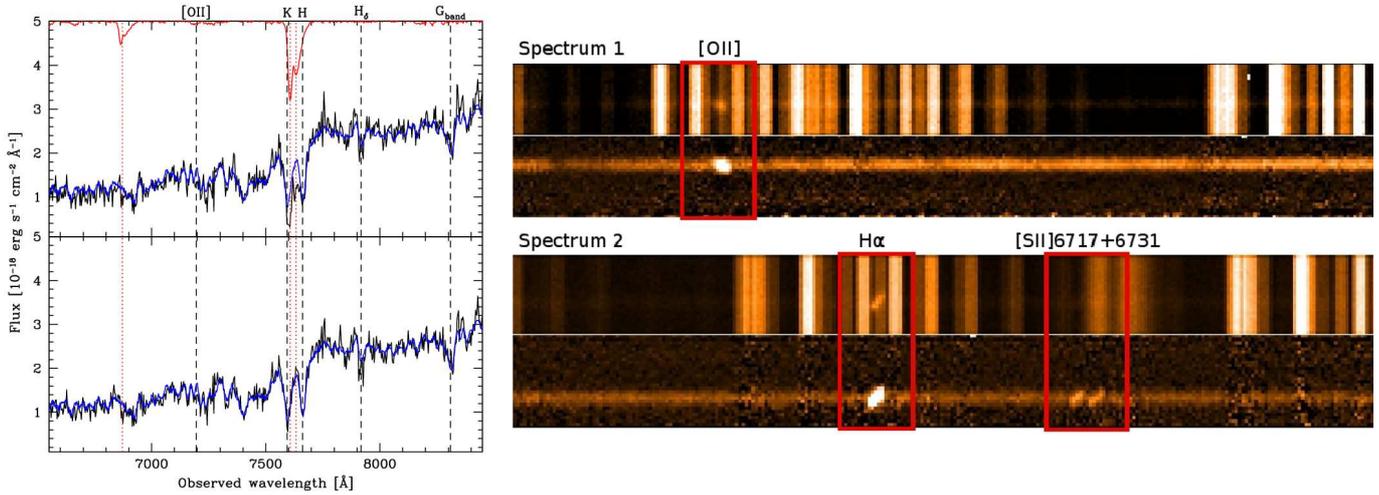}
\caption{Example of the good quality achievable through the removal of absorption (left) and emission (right) sky features with \textit{F}-VIPGI. \textbf{\textit{Left panel:}} Atmospheric correction for a spectrum of a passive galaxy at $z \sim 0.9$ observed with grism 300I. \textit{Top:} Flux-calibrated spectrum (in black) without the atmospheric correction applied. It is evident that the telluric lines at $\lambda \sim$ 7600\AA{} strongly affect the S/N of the 4000-\AA{} break and the CaII (H, K) lines.
\textit{Bottom:} The same spectrum after applying the atmospheric absorption correction. Now the CaII lines are completely recovered. In red (solid and dotted curves) the positions of the most prominent telluric lines are shown. The black dashed lines mark the position of the most important spectral features in the displayed wavelength range. The best-fitting template of both spectra (an LRG one at $z$ = 0.930) is overlaid in blue in both panels for reference. \newline
\textbf{\textit{Right panel:}} Subtraction of the emission sky lines from two spectra with different redshift (\textit{spectrum1} at $z$ = 1.086, \textit{spectrum2} at $z$ = 0.282). For each object the same part of the spectrum before and after the sky subtraction is shown at the top and bottom, respectively. In addition, the strongest spectroscopic features of each object are marked by the red boxes and are labeled on the top side. Although some residuals of the sky removal are visible in the bottom panels, the pipeline efficiently subtracts the undesired sky features and recovers the science features, which are sometimes completely outshone by the atmospheric emission.
}
\label{fig:SkyCorrection}
\end{figure*}

A significant source of noise along the spectroscopic reduction process is introduced by the strong absorption (telluric) and emission sky features produced by the O$_2$ and H$_2$O molecules and the OH$^-$ radical at $\lambda >$ 6000\AA{} (see Fig.\,\ref{fig:SkyTell}, left). \textit{F}-VIPGI is able to remove the two undesired sky contributions in a very efficient way, as shown in Fig.\,\ref{fig:SkyCorrection}.
The sky emission spectrum is computed for each slit in the \textit{free regions}\footnote{These regions are identified automatically by the pipeline with the following procedure: first a ``slit profile'' is traced by collapsing the data along the wavelength axis. The mean signal level and rms (sigma) of this slit profile are then computed with a robust iterative procedure by using the biweigth estimator described by \cite{Beers1990a}. Finally, all groups of at least three pixels that are \textit{N} sigma above the mean level are considered as object spectra while everything else is considered free sky region. In the above process the sigma threshold (\textit{N}) can be set by the user.} on the sides of the science spectrum. If this area is large enough (at least 2\arcsec\ left between the spectrum and the slit edges, on both sides), the sky lines can be accurately modeled and efficiently subtracted from the final spectrum, as shown in the right panels of Fig.\,\ref{fig:SkyTell} and \ref{fig:SkyCorrection}. After sky subtraction, we need to correct all the spectra also for the prominent sky absorption features (telluric lines) that could otherwise be interpreted as absorption lines in the object spectra. These spurious features can be removed provided that a sufficient number of well exposed flux calibrated spectra are available. Even if the best results are obtained when the spectra selected for the correction span a wide redshift range (because the intrinsic features are efficiently removed in this way and only the sky ones are enhanced), a good telluric line absorption correction can also be gained by using cluster galaxies, whose redshift values are relatively close to each other. If so, the pipeline finds the atmospheric correction by combining the spectra whose telluric lines are more evident and isolated and finally applies such a correction to \textit{all} slits of the mask. An example of the good quality achievable for telluric line absorption correction is shown in the left panel of Fig.\,\ref{fig:SkyCorrection}.

\subsection{The final product}
\label{par:final_product}
Once the preliminary reduction and the wavelength calibration are applied to each frame, these are calibrated in flux (see Sect.\,\ref{par:SPHcalib}) and finally combined to obtain a stack of 2D sky-subtracted spectra. Single spectra are then extracted from the 2D stacked frame, using a Horne optimal extraction \citep{Horne1986}, and are saved in local folders in fits format. For each spectrum \textit{F}-VIPGI also produces the associated noise vector generated by the extraction residual.
\begin{figure*}[ht]
\sidecaption
\includegraphics[height=5cm, clip=true]{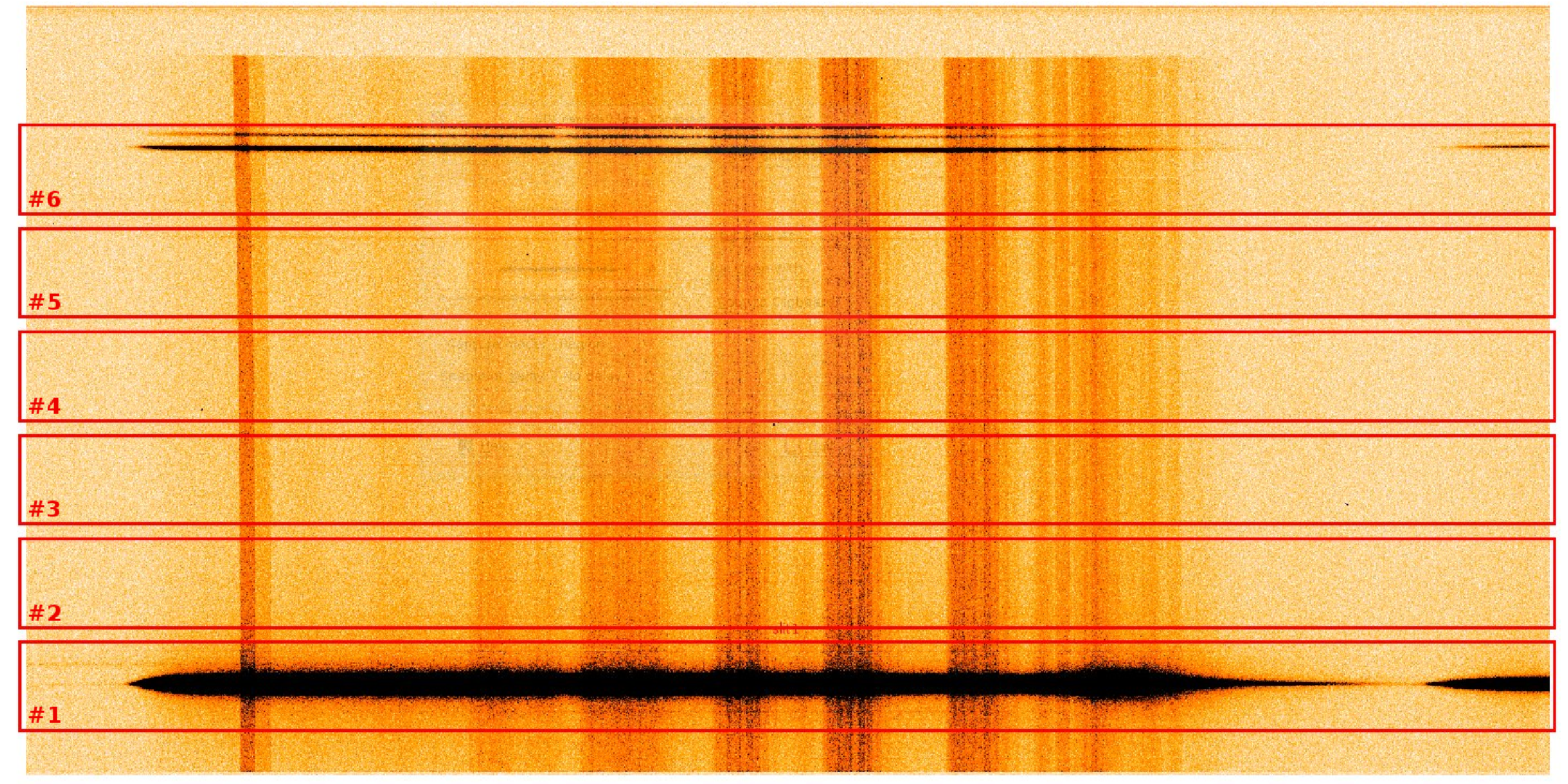} 
\caption{Image of the spectrum of a standard star taken with the FORS2 300I instrument setup in the first chip (Q1). In red are shown the \textit{virtual} slits of 30\arcsec\ length that \textit{F}-VIPGI creates in the header of the scientific and calibration frames of the standard stars to mimic the VIMOS setting. In FORS2 observations the standard star spectra are always imaged in Q1 and are always enclosed in the first (\#1) virtual slit.}
\label{fig:STDslitQ1}
\centering \vspace{0.5cm}
\includegraphics[height=5.9cm, clip=true]{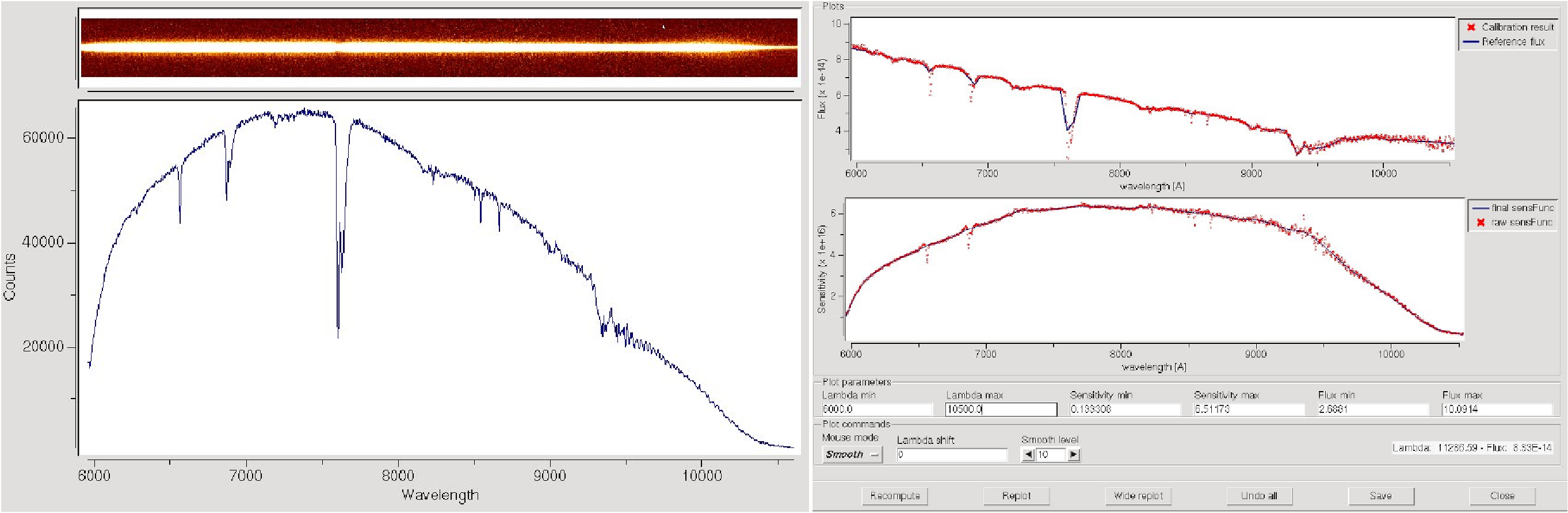}
\caption{Example of the spectrophotometric calibration for a FORS2 observation with grism 300I and sorting filter OG590. \textbf{\textit{Left panel}}: Extracted 1-D spectrum of the standard star, not yet corrected for the CCD response function.
\textbf{\textit{Right panel}}, \textit{top}: The expected spectrum of the standard star provided by the SPHOT table (continuous line) and the fitted one (red points) according to the spectral resolution of the observation. The sensitivity function (\textit{bottom panel}) is then computed as the ratio between the extracted 1-D spectrum and the fitted standard star spectrum converted into erg\,cm$^{-2}$\,s$^{-1}$\,\AA{}$^{-1}$. The user can edit the originally computed sensitivity function (red points) and choose the final one (continuous line) to apply to the single spectra for calibrating them into flux.
}
\label{fig:SPHcalibration}
\end{figure*}
\subsection{Spectrophotometric calibration}
\label{par:SPHcalib}
An important difference of the FORS2 observations to those of VIMOS concerns the methods used by the two instruments to acquire the spectra of spectro-photometric standard stars (STD). This step is mandatory to calibrate each spectrum in flux and to correct its continuum shape for the CCD sensitivity function.
Standard stars in VIMOS are observed in all four quadrants and with a specifically designed mask composed of eight slits per chip, with fixed position and length (10\arcsec). In FORS2, instead, the STD frame is acquired only on Q1 and by means of a single \textit{long-slit} with 5\arcsec\ width.

To overcome this discrepancy and enable the use of the standard VIPGI routines for spectrophotometric calibration also for FORS2 data, \textit{F}-VIPGI writes some keywords into the header of the standard star frames (and into the correlated calibration files) that \textit{mimic} the presence of 11 slits in total (6 in Q1 and 5 in Q2)\footnote{This is the maximum number of 30\arcsec\ length slits that can be uniformly placed on the area of the two chips.} with 30\arcsec\ length and 1\arcsec\ width. It is important to note that in FORS2 observations the standard star spectra are always centered on Q1 at the position given by the header keywords \texttt{CRPIX1} and \texttt{CRPIX2}. The virtual slits are therefore created in the way to have the first slit (\#1 of Fig.\,\ref{fig:STDslitQ1}) always centered on (\texttt{CRPIX1}, \texttt{CRPIX2}) and, so, always contain the star spectrum. An example of the part of the \textit{virtual} mask relative to Q1 is shown in Fig.\,\ref{fig:STDslitQ1}.

Once the standard-star spectrum is reduced and extracted, the pipeline computes the sensitivity function by comparing the observed stellar spectrum with the expected one contained in the SPHOT table (Sect.\,\ref{par:fors2Calibration}) as shown in Fig.\,\ref{fig:SPHcalibration}. The extracted curve is finally saved in a calibration table that can be applied to the 1-D spectra for their flux calibration and for the correction of the CCD response function. Since standard stars in FORS2 are observed only on Q1, for Q2 data the same calibration table as computed for the first chip has to be used.
An example of how the sensitivity function is computed for a standard star observed with the 300I grism is shown in Fig.\,\ref{fig:SPHcalibration}.

\subsection{Redshift accuracy test}
\label{par:selfCons_test}
The XDCP massive cluster XMMUJ1230.3+1339 at $z$ = 0.975 \citep{Fassbender2011a0,Lerchster2011} has been targeted by four different FORS2 spectroscopic observations in order to fully characterize its extended and dynamically unrelaxed galaxy population. This resulted in 65 confirmed spectroscopic members, 15 of which had multiple (double) observations. This enabled us to use these targets for a self-consistency test on the \textit{F}-VIPGI redshift measurements, based on different independent spectroscopic observations of the same object.

To this aim we selected 14 out of the 15 duplicated observations, considering only those couples where the final extracted data were good enough to provide a safe redshift assessment for both cases.
The median of the differences (in absolute value) between the redshift measurements is $<|\Delta z|> = 5.1\cdot10^{-4}$, with a spread given by the semi-interquartile range (SIQ) of $<|\Delta z|>_{SIQ} = 2.5\cdot10^{-4}$. This value can be translated into a corresponding \textit{rest-frame} velocity uncertainty of $\delta_{v,rest} \approx$ 77 km s$^{-1}$ for a galaxy at $z$ = 0.975.

\begin{figure}[t]
\centering
\includegraphics[height=9cm, clip=true]{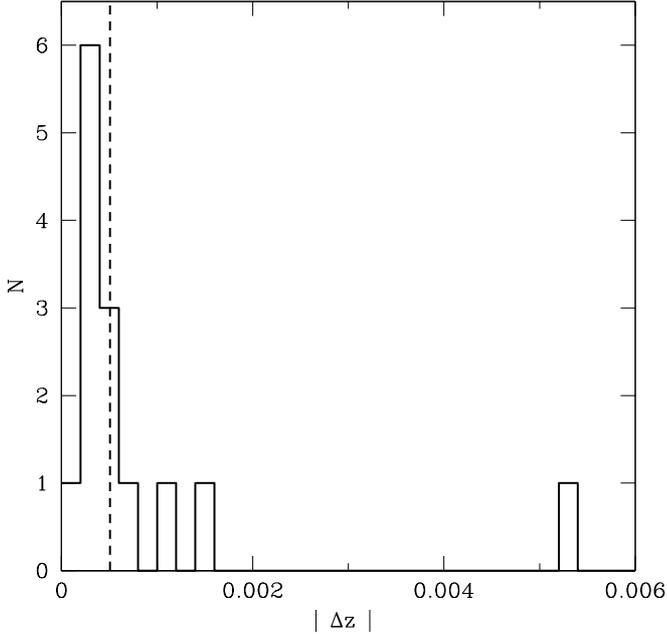} 
\caption{Distribution of the absolute values of the differences between the redshift measurements $|\Delta z|$ for 14 members of XMMUJ1230.3+1339 at $z$ = 0.975 from multiple, independent, spectroscopic observations. The median value of $<|\Delta z|>$ = 5.1$\cdot10^{-4}$ is marked by the vertical dashed line.\vspace{-0.3cm}}
\label{fig:delta_zHist}
\end{figure}

The distribution of $|\Delta z|$ is shown in the histogram of Fig.\,\ref{fig:delta_zHist}.
\section{An application to a sample of distant galaxy clusters}
\label{par:application}
In this section we show an application of \textit{F}-VIPGI to the analysis of spectroscopic data for a sample of 16 distant XDCP galaxy clusters in the redshift range (0.65 $\leq z \leq$ 1.25) observed with the grism 300I+11 (Table\,\ref{Tab:fors2GRISM}).
The main goal of this section is to provide the community with a new set of spectroscopic templates that are intended to represent the galaxy population residing in distant galaxy clusters. For this reason, as shown in Sect.\,\ref{par:passive_spec_comparison}, the present library is completely different from the existing ones (e.g., the ones constructed from SDSS, zCOSMOS or K20 surveys), which contain templates representative of the population of field galaxies at different redshifts up to $z \sim$ 1.5. In addition to identifying redshifts of distant galaxies, the provided templates can also significantly improve the efficiency and the reliability of the \textit{photometric} redshift assessment of distant clusters with complex star formation histories (see, e.g., \cite{Guennou2010} and \cite{Pierini2012} for a discussion) and they can be used to predict rest-frame $U-B$ colors of cluster galaxies at high redshifts.

\subsection{The spectroscopic sample}
\label{par:dataSample}
\begin{figure}[t]
\centering
\includegraphics[height=9cm, clip=true]{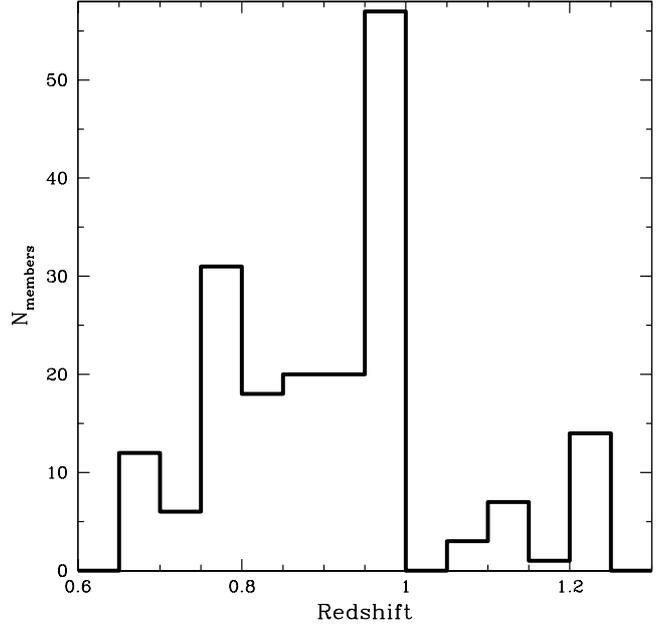} 
\caption{Redshift distribution of the selected 187 galaxies, members of the 16 XDCP clusters reported in Table\,\ref{Tab:clusterList}.\vspace{-0.2cm}}
\label{fig:zDistributionGals}
\end{figure}

For our study we selected all the XDCP clusters with at least three spectroscopically confirmed members with an S/N $>$ 2 in the rest frame wavelength range of 4000 - 4300 \AA{} in their reduced spectra. Each spectrum was also visually inspected to verify that no contaminations from cosmic rays or bad sky subtraction were present. 
For the cluster member selection, we firstly estimated the redshift cluster ($z_{cl}$) as the median of the redshift peak found in the proximity of the X-ray emission center. After that, we selected the cluster members as those galaxies with a rest-frame velocity offset $<$\,3000\,km\,s$^{-1}$ from $z_{cl}$, corresponding to a redshift window cut of $\Delta z < 0.01 \times (1+z_{cl})$.
\begin{table*}[ht]
\centering          
\caption{List of the 16 XDCP clusters used for our spectroscopic analysis.}
\begin{tabular}{c c c c}
\vspace{-0.3cm}            \\   
\hline\hline    \vspace{-0.3cm}            \\   
ID & $z_{spec}$& N$_{members}$ & References\\
\vspace{-0.3cm}            \\ 
\hline \vspace{-0.3cm}            \\
cl1 	& 0.7690 & 13 & (1) \\  
cl2 	& 1.2310 & 8  & (1)\\
cl3 	& 0.9410 & 16 & (1)\\
cl4	& 0.9590 & 6  & C20 in (2)\\
cl5	& 0.8290 & 8 & (1)\\
cl6	& 0.7850 & 5  & (1)\\  
cl7	& 0.6770 & 12 & (1)\\
cl8	& 0.8830 & 4  & (1)\\
cl9	& 0.8940 & 19 & (1)\\
cl10	& 0.7890 & 13 & (1)\\
cl11	& 1.1220 & 7 & C11 in (2)\\   
cl12	& 1.0740 & 3  & (1)\\
cl13	& 0.8270 & 9 & (1)\\
cl14	& 0.7460 & 7  & (1)\\
cl15	& 0.9750 & 52 & XMMU J1230.3+1339 in (3) and (4)\\
cl16	& 1.2030 & 7  & (1)\\
\vspace{-0.3cm}            \\
\hline \hline   \\
\end{tabular}
\tablebib{(1) To be published; (2) \cite{Fassbender2011b}; (3) \cite{Fassbender2011a0}; (4) \cite{Lerchster2011}}
\label{Tab:clusterList} 
\end{table*}
The list of the final targets selected for the next spectroscopic analysis is reported in Table\,\ref{Tab:clusterList}. For each target a sequential ID, the spectroscopic redshifts, the number of confirmed members used in the subsequent analysis, and the literature reference (if existing) are reported.
A total of 187 cluster members have been identified in this way, with a redshift distribution in the range 0.65 $\leq z \leq$ 1.25 as shown in Fig.\,\ref{fig:zDistributionGals}.

\subsection{Spectral indices analysis results}
\label{par:specIndex}
\begin{table*}[ht]
\centering          
\caption{Classification criteria taken from \cite{Poggianti2009} used in this study for grouping the cluster galaxies into five pectral types according to the equivalent widths values of their [OII]$\lambda$3727 and H$\delta$ lines. The following convention is used: $EW < 0$ for \textit{emission} lines, $EW > 0$ for \textit{absorption} lines. The number of galaxies found in each class is given in the last column.}
\begin{tabular}{l c c c}
 \vspace{-0.3cm}            \\   
\hline\hline       \vspace{-0.3cm}            \\
Spectroscopic class 		& EW([OII]) value 	& EW(H$\delta _A$) value & Number \\ 
				&	[\AA{}]		&	[\AA{}]		 & of galaxies\\
\vspace{-0.3cm}            \\
\hline                    
\vspace{-0.3cm}            \\
Passive 			& $>$-5 		& $<$3			 & 108\\  
Post-starburst		 	& $>$-5 		& $\geq$3		 & 21\\
Quiescent star-forming  	& ]-25, -5]	 	& $<$4			 & 23\\
Dusty starburst			& ]-25, -5] 		& $\geq$4		 & 14\\
Starburst  			& $\leq$-25 		& any 			 & 21\\
\vspace{-0.3cm}            \\
\hline \hline
\end{tabular}
\label{Tab:specClassification} 
\end{table*}

We grouped the above galaxies into five different spectral classes according to the measured\footnote{All spectral indices were measured in an automated way by means of a Python script developed by our group that uses the trapezoidal rule to perform the integration. The code functioning was tested on some spectra by comparing its results with those provided by the standard IRAF packages.} values of the equivalent widths (EW) of their [OII] and H$\delta$ lines, as summarized in Table\,\ref{Tab:specClassification}. These two spectral features are, in fact, reliable indicators of ongoing and recent star formation activities within timescales of $\sim 10^7$ and $\sim 10^9$\,yr for [OII] and H$\delta$, respectively. The amplitude of their 4000-\AA{} break (D4000) and the equivalent widths of $H\beta$, [OIII]$\lambda4959$ and [OIII]$\lambda5007$ were also computed but were not used for the spectral classification. 
The errors on the indices were estimated by means of the noise vector provided by \textit{F}-VIPGI for each extracted spectrum at the end of the reduction process (see Sect.\,\ref{par:final_product}). Specifically, for each spectrum 1000 Monte-Carlo realizations were generated using the related noise vector as variance. The uncertainties on the indices were then computed as the rms of the results obtained by repeating the spectral indices analysis on the series of simulated spectra.

We stress that the relative fraction of galaxies in the different spectral types is substantially biased by the method used in XDCP to select the targets for the spectroscopic follow-up. As discussed in \cite{Fassbender2011b}, to maximize the probability of targeting actual cluster members, FORS2 masks are created in a fashion that the slits are preferentially placed on color-selected galaxies close to the expected red-sequence color and within the detected X-ray emission. This implies that the \textit{majority} of the observed targets is actually expected to be passive and, hence, that no statistically significant conclusions about the relative number of galaxies in the different classes can be inferred from the last column of Table\,\ref{Tab:specClassification}.

We adopted the bandpasses defined in \cite{Balogh1999} for EW([OII]) and D4000 (hereafter D$_n$4000) while for EW(H$\delta$) we used the definition of H$\delta _A$ given by \cite{Worthey1997}. The galaxies were classified into five spectral classes following the same method as described in \cite{Poggianti2009}, which is based only on the strength of [OII] and H$\delta _A$ lines, as summarized in Table\,\ref{Tab:specClassification}.

\begin{figure}[ht]
\centering
\includegraphics[height=9cm, clip=true]{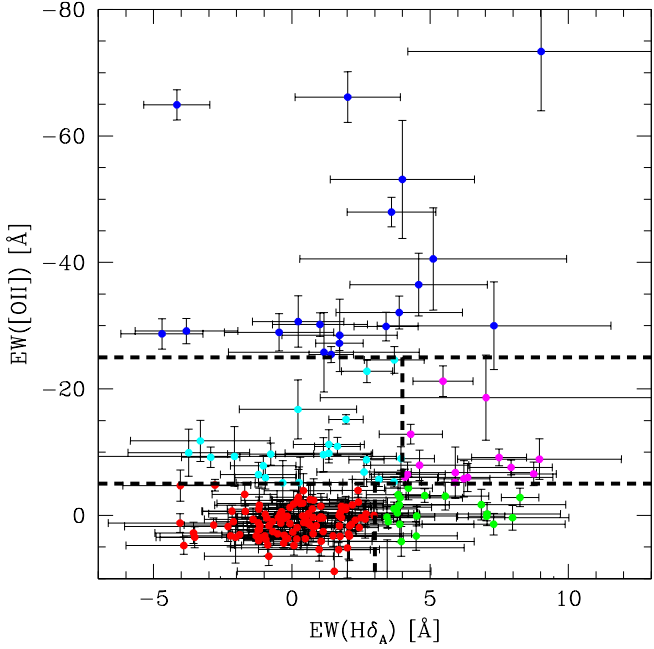} 
\caption{Distribution of the 187 galaxies in EW([OII]) - EW(H$\delta _A$) plane. The dashed lines and the different colors mark the positions of the five spectral classes, with the same criteria as in \cite{Poggianti2009}. The color code is the following: \textit{blue}: starburst galaxies; \textit{cyan}: quiescent star-forming; \textit{magenta}: dusty-starburst; \textit{red}: passive; \textit{green}: post-starburst.
}
\label{fig:OII_Hdelta}
\end{figure}

Fig.\,\ref{fig:OII_Hdelta} shows the distribution of the 187 galaxies in the EW([OII]) - EW(H$\delta _A$) plane. The adopted loci for the five spectral classes are marked by the dashed lines and a color coding was used for a better visualization of the groups.

\begin{figure}[ht]
\centering
\includegraphics[height=8.5cm, clip=true]{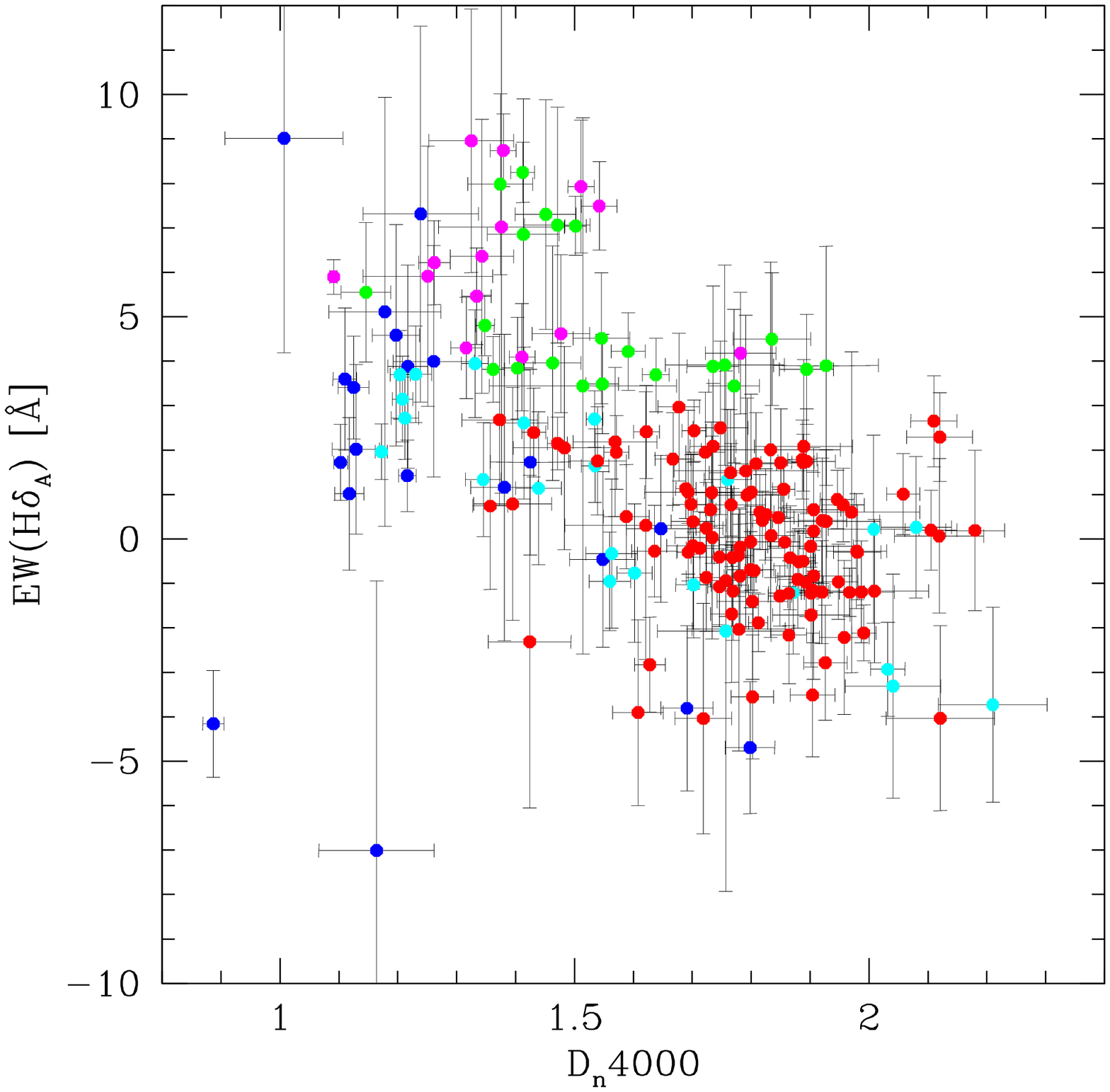}
\includegraphics[height=8.5cm, clip=true]{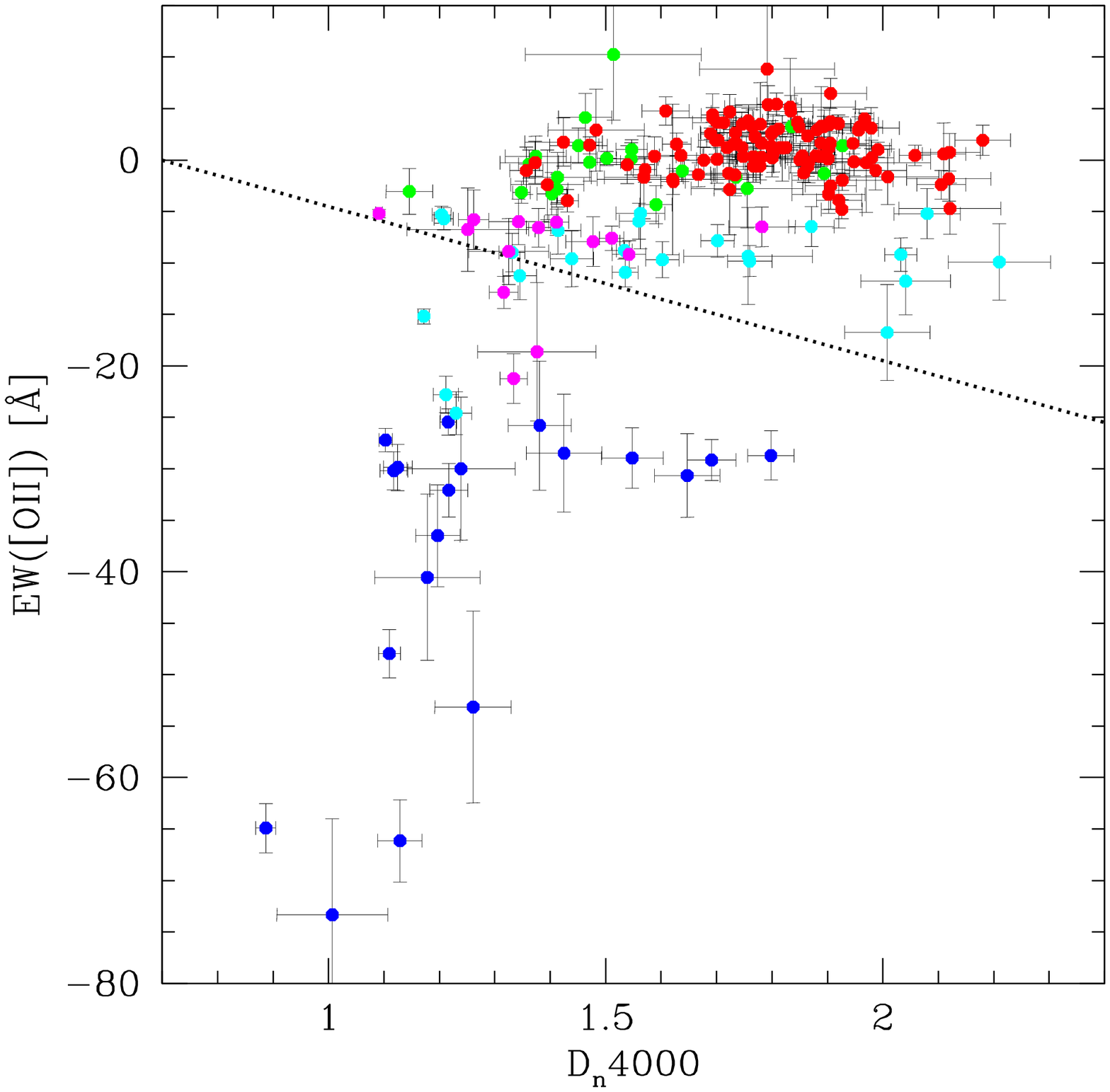}  
\caption{D$_n$4000 versus EW([OII]) (\textit{top} panel) and EW(H$\delta _A$) (\textit{bottom} panel) for the selected targets. The dotted line in the right panel represents the division between early-type / late-type galaxies adopted in \cite{Franzetti2007}. The color code is the same as in Fig.\,\ref{fig:OII_Hdelta}.
}
\label{fig:d4000_HdA_OII}
\end{figure}

The D$_n$4000 index was then used to qualitatively test the validity of our classification. The top panel of Fig.\,\ref{fig:d4000_HdA_OII} shows the distribution of our galaxy sample in the EW(H$\delta _A$) vs D$_n$4000 space with the color code adopted in Fig.\,\ref{fig:d4000_HdA_OII} for identifying the different spectral classes. The distribution we find is remarkably similar to that reported in Fig.1 of \cite{Gallazzi2009} for a set of model galaxies. Namely, we observe that consistently with the results of these authors, the starburst (dusty and not) galaxies tend to be concentrated on the upper left side of the plot, indicating a relatively young (and still forming) stellar population. The passive and quiescent star-forming galaxies instead reside mostly in the center and on the lower right side, suggesting that they experienced the last episode of major star-formation activity more than 3 Gyr earlier. Finally, the \textit{post-starbursts} are all located above the sequence drawn by the passive and quiescent star-forming galaxies, a sign that their starburst activity ended within the previous 2 Gyr.

In the bottom panel of Fig.\,\ref{fig:d4000_HdA_OII} the distribution in the D$_n$4000 - EW([OII]) diagram is shown. It is evident also here that for the adopted classification method the different spectral classes are confined to specific regions of the plots. In particular if we consider the relation adopted by \cite{Franzetti2007},
 \[
\mbox{D}_n4000 + EW(\mbox{[OII]})/15 > 0.7                                                                                                       
\]
(marked by the dotted line in the plot) to divide passively evolving (\textit{early-type}) galaxies from star-forming (\textit{late-type}) objects for a spectroscopic sample of VVDS galaxies at 0.45\ $\leq z \leq$\ 1.2, all passive and post-starburst galaxies appear to be efficiently separated from the starburst ones.
However, the plot also shows that the majority of quiescent star-forming and dusty-starburst galaxies are classified as early-type in this way. This effect was already predicted by the same authors, who claimed that some contamination in the early-type region was expected due to early spirals and that this is a general feature observed for color- or spectroscopic classification schemes.

To identify possible active galactic nuclei (AGNs, e.g. Seyfert 2) erroneously identified as starburst galaxies, we used the diagnostic diagram of \cite{Bongiorno2010b}, that was originally defined in \cite{Lamareille2004b} (and references therein), where the ratios between the EW of [OII]$\lambda$3727, H$\beta$, and [OIII]$\lambda$5007 lines are used to separate star-forming from AGN-photoionized spectra. The constraint of having the [OIII]$\lambda$5007 line within the observed wavelength range 6000\AA{} $< \lambda _{obs} <$ 9800\AA{} makes this analysis feasible only for seven starburst galaxies lying at redshift $z <$ 0.95.

\begin{figure}[ht]
\centering
\includegraphics[height=8cm, clip=true]{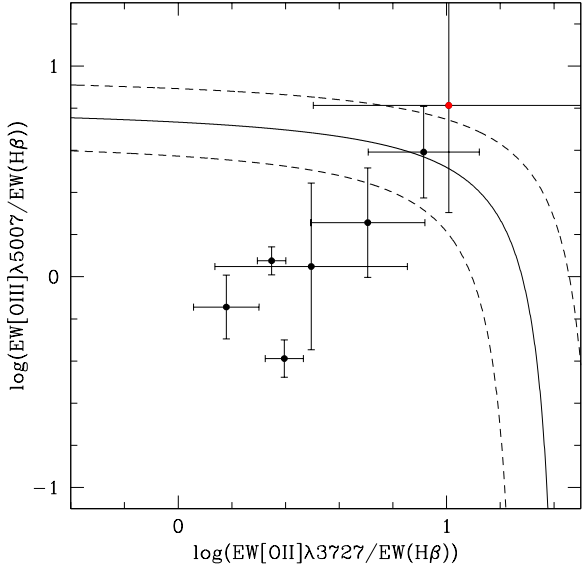} 
\caption{Diagnostic diagram used by \cite{Bongiorno2010b} to distinguish between pure star-forming and AGN-contaminated galaxies applied to those galaxies of our sample classified as starburst (see Table\,\ref{Tab:specClassification}) and lying at redshift 0.6\,$< z <$\,0.95 to have their [OIII]$\lambda$5007 line falling within the observed wavelength range. The solid and dashed curves show the demarcation and its $\pm0.15 dex$ uncertainty, respectively, between pure star-forming galaxies (bottom region) and AGN (top region), as defined in \cite{Lamareille2004b}. The red point marks the starburst object possibly hosting an AGN that was therefore excluded from the final stacking procedure.}
\label{fig:diagnosticAGN}
\end{figure}

In Fig.\,\ref{fig:diagnosticAGN} the distribution of these objects in the log(EW[OII]$\lambda$3727/EW(H$\beta$)) vs log(EW[OIII]$\lambda$5007/EW(H$\beta$)) plane is shown. Only one object (marked in red) appears to lie in the ``AGN region" of the plot and was consequently excluded from the final stacking of the starburst spectra.

\begin{table*}[ht]
\centering          
\caption{Spectral properties of the five averaged spectra of Fig.\,\ref{fig:stackSpectra}. The reported S/N values refer to the wavelength range 4000\AA{}\,$<$\,$\lambda$\,$<$\,4300\AA{}.}
\begin{tabular}{l c c c c c c}
    \hline\hline       \vspace{-0.3cm}            \\
    Spectroscopic class 	&S/N	& EW([OII]) 	   &EW(H$\delta _A$)&   D$_n$4000  &EW(H$_{\beta}$) &EW([OIII]$\lambda$5007)\\ 
			  &		&[\AA{}]	   &[\AA{}]	    &	           &[\AA{}]	    &[\AA{}]		     \\
    \vspace{-0.3cm}            \\
    \hline      
    \vspace{-0.3cm} \\ 
    Passive 		&9	&    1.431$\pm$1.191& 0.148$\pm$0.981& 1.810$\pm$0.030 &  1.829$\pm$0.606  &  0.958$\pm$0.611\\  
    Post-starburst		&6 	&   -0.178$\pm$1.739& 4.910$\pm$1.395& 1.543$\pm$0.034 &  2.753$\pm$1.164  &  0.540$\pm$1.448\\
    Quiescent star-forming  &7 	&   -7.946$\pm$1.791& 0.593$\pm$1.231& 1.550$\pm$0.031 &  1.664$\pm$0.887  & -1.184$\pm$0.985\\
    Dusty starburst		&7.5	&   -7.719$\pm$1.605& 5.827$\pm$1.123& 1.412$\pm$0.030 & -2.357$\pm$1.849  & -5.842$\pm$0.693\\
    Starburst  		&5	&  -34.943$\pm$2.929& 1.368$\pm$1.887& 1.254$\pm$0.036 & -8.816$\pm$1.797  &-13.516$\pm$2.256\\
    \vspace{-0.3cm}            \\
    \hline \hline    \\              
    \end{tabular}
\label{Tab:specIndex_Templates} 
\end{table*}

\section{A new library of spectroscopic templates}
\label{par:library}
Before proceeding with the final stacking procedure, we restricted the usable portion of each spectrum to the observed wavelength range of 6100 - 9280 \AA{}. This cut was made to avoid the higly sky-contaminated region at $\lambda \geq$ 9300\AA{} (Fig.\,\ref{fig:SkyTell}, left) while simultaneously preserving the observability of spectral features like $H\beta$ and [OIII] for galaxies up to $z \sim$ 0.85.

\begin{figure}[htb]
\centering
\includegraphics[width=9.5cm, clip=true]{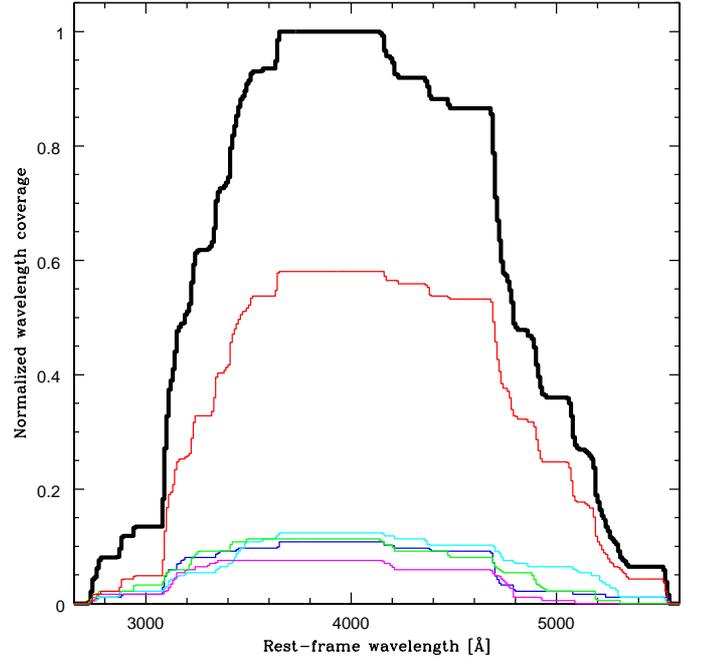} 
\caption{Normalized rest-frame wavelength coverage for the 186 galaxies used in the stacking procedure. The general result for all galaxies is represented by the thick solid curve, while the thinner lines indicate the different contributions from the five spectral classes defined in Sect.\,\ref{par:specIndex}. The same color code as in Fig.\,\ref{fig:OII_Hdelta} is used.}
\label{fig:lambda_coverage}
\end{figure}

The total rest-frame wavelength range covered by the composite spectra is 2700 - 5300 \AA{}, with a normalized \textit{wavelength coverage} for all 186 considered galaxies shown in Fig.\,\ref{fig:lambda_coverage}. In the same figure we also show the different contributions from the five spectral classes defined in Sect.\,\ref{par:specIndex}. We recall that here the possible AGN-contaminated starburst object of Fig.\,\ref{fig:diagnosticAGN} was excluded from our analysis.

For the final stacking process each spectrum was rescaled to the mean flux computed in the wavelength range 4000\AA{} $< \lambda <$ 4300\AA{} and then combined with the other objects of the same spectral class using the median as operator.

\begin{figure*}[ht]
\centering
\includegraphics[width=17.9cm, clip=true]{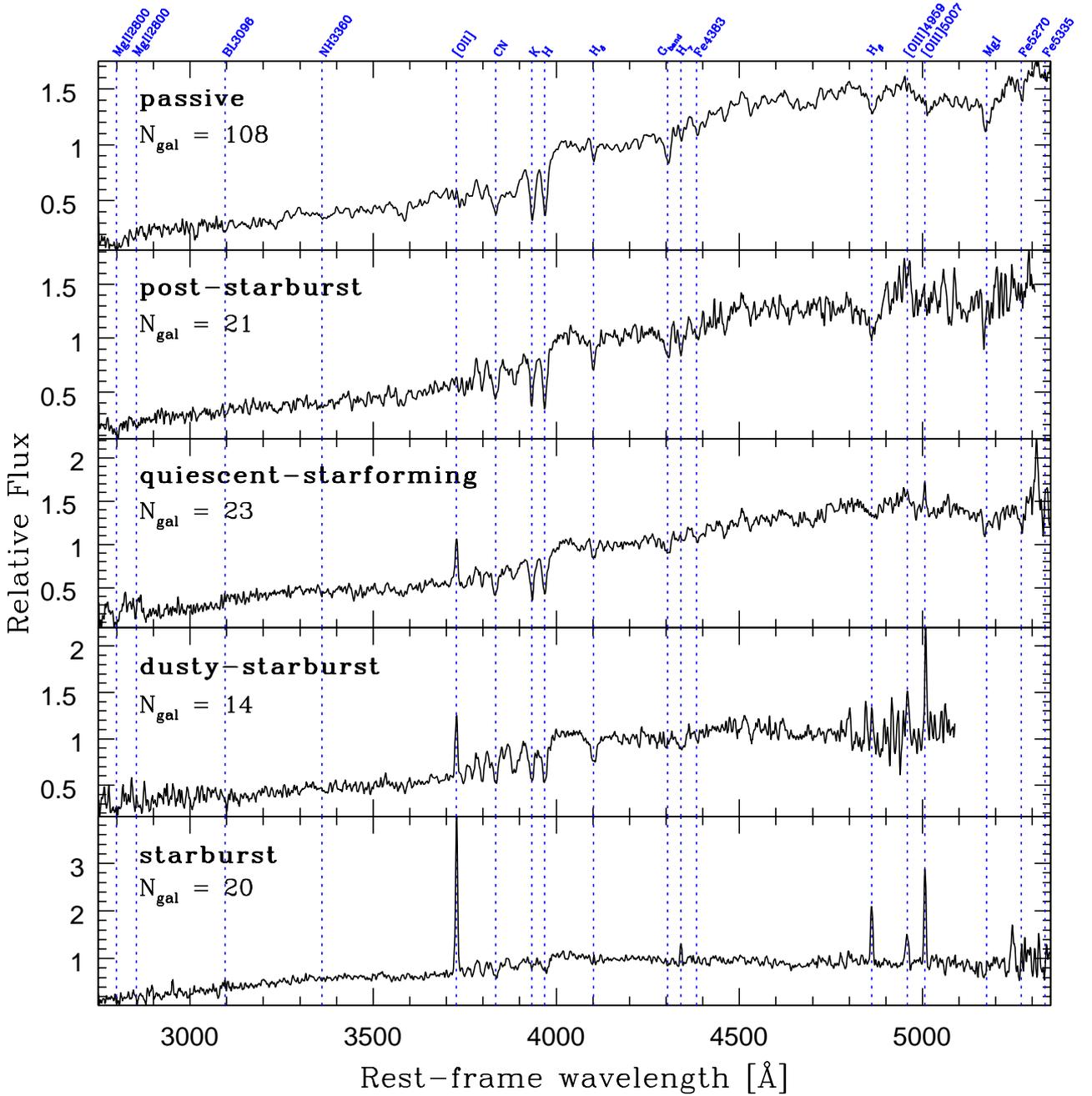} 
\caption{Final composite spectra, representative of the five considered spectral classes, created from a total of 186 galaxy spectra. The spectral classes and the number of stacked spectra are reported in the top left corner of each panel together with the expected position of the main spectral features marked by the blue vertical lines. A Gaussian smoothing filter of three pixels was applied.
}
\label{fig:stackSpectra}
\end{figure*}

The resulting five composite spectra are shown in Fig.\,\ref{fig:stackSpectra}. 
For each spectrum the relative spectral class and the number of galaxies used in the stacking process are reported in the top left corner. The most important spectral features recognizable in the rest-frame wavelength range $\Delta \lambda$ = 2700 - 5300 \AA{} are labeled on the top side of the plot, and their expected positions are marked by dashed vertical lines.

The spectral properties of the average spectra are quantified in Table\,\ref{Tab:specIndex_Templates} together with their S/N computed in the wavelength range 4000\AA{} $< \lambda <$ 4300\AA{}.

We note that the values of EW(H$\delta _A$) and, especially, D$_n$4000 obtained for the ``passive'' template are similar to those observed for the most massive and passive galaxies in the local universe by SDSS-DR4 (D$_n$4000$|_{z = 0}$ $\approx$ 1.9; EW(H$\delta _A$)$|_{z = 0}$ $\approx$ -1.5) that were reported e.g. by \cite{Gallazzi2005}.

In the next two sections (\ref{par:passive_spec_comparison} and \ref{par:U_B_Coma_comparison}) we compare the spectro-photometric properties of our composite spectra with those of galaxies residing in different redshift ranges and environments.

\subsection{Comparison of the resulting passive template with previous library spectra}
\label{par:passive_spec_comparison}
To highlight the original aspects of the spectroscopic library provided in this paper and, hence, its uniqueness w.r.t. other previous similar works, we considered as an example the composite spectrum of our passive galaxies and qualitatively compared it with others representative of the same class of objects that reside in the local and distant universe and in different environments.
Namely, we considered the distant sample of \textit{early-type} galaxies in the field studied within the K20 survey, the composite spectrum of which has been provided by \cite{Mignoli2005}. We highlight that the number of passive (stacked) spectra (93), their typical redshift values ($< z >$ $\sim$\,1.0, extended up to $z\,\sim$\,1.25) and the spectro-photometric method used by the authors to identify their passive galaxies are very close to the numbers and criteria adopted in our work (see Sect.\,\ref{par:application}).
As a second reference sample we considered the nearby passive template provided by \cite{Kinney1996}, produced from a total of four morphologically selected bright (M$_B <$\,-21) elliptical galaxies residing in groups with $< z > \sim$\,0.008.
Finally, we adopted the composite spectrum produced by \cite{Eisenstein2003} by using more than 1000 SDSS luminous red galaxies (LRG), selected in magnitude (M$_g <$\,-21.8) and in the redshift range 0.3\,$< z <$\,0.55, as a third element of comparison.
\begin{figure*}[htb]
\centering
\includegraphics[width=18cm, clip=true]{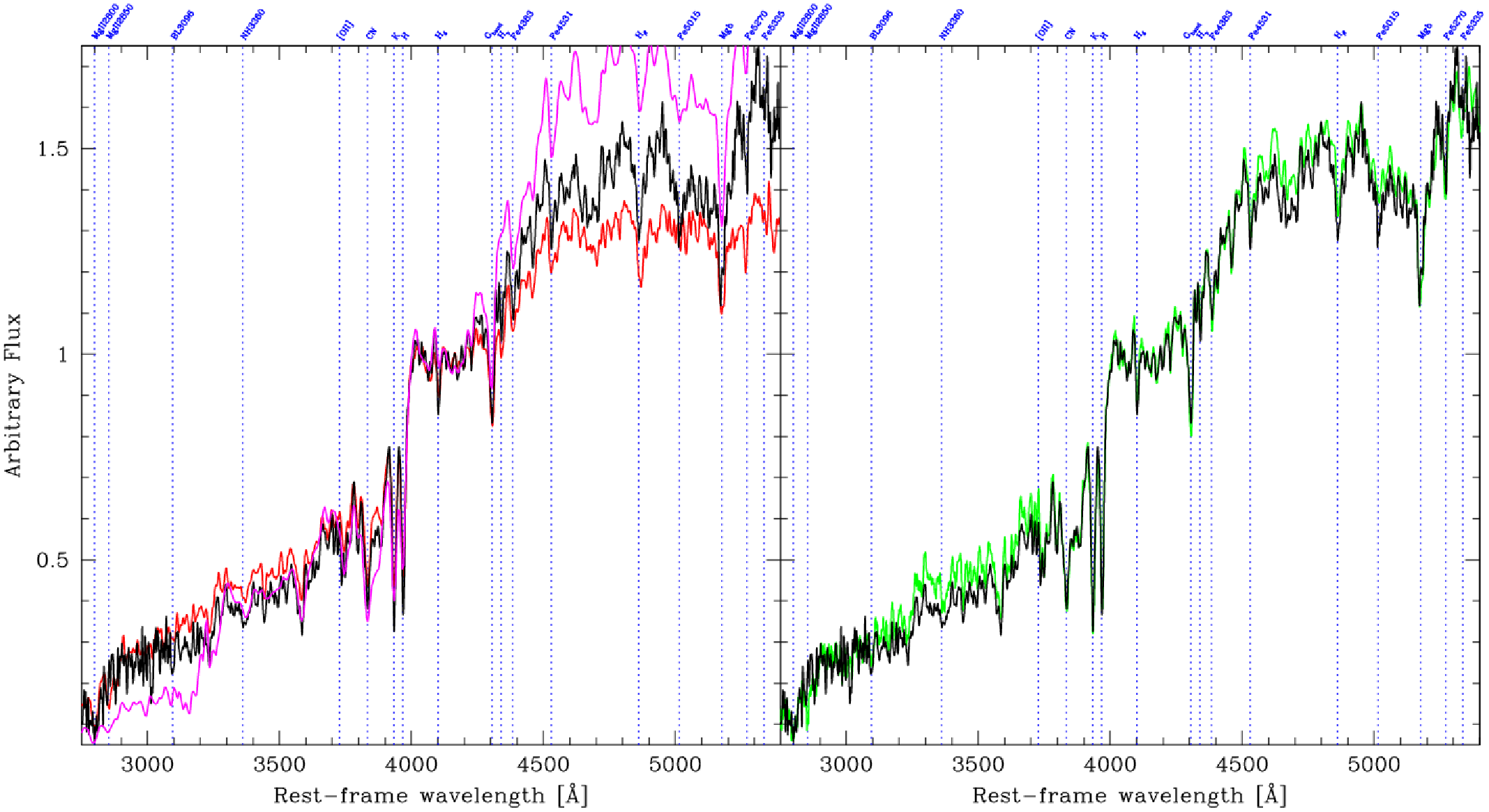}
\caption{Qualitative comparison of our composite passive spectrum (black line) with the empirical templates quoted in the text that is representative of passive galaxies residing in different environments and redshift ranges. All spectra are normalized in the wavelength range 3950\AA{} $< \lambda <$ 4050\AA{} and smoothed with a three pixel boxcar filter.
\textit{Left:} The K20 composite spectrum of distant early-type galaxies in the field is shown in red while the template computed by \cite{Kinney1996} for elliptical galaxies in groups at z\,$\sim$\,0.008 is shown in magenta. It is evident that the passive galaxies of our sample are \textit{redder} w.r.t. the K20 ones at similar redshift, but bluer than the ellipticals in groups in the local universe.
\textit{Right:} Comparison between the average spectrum of our passive galaxies and the LRG template provided by \cite{Eisenstein2003} (in green) that was computed for $z\,\sim$\,0.4, M$_g < -21.8$ galaxies. Although the two spectra were obtained by using different galaxy selection criteria and redshift ranges, they appear to be remarkably similar overall (except for a slightly higher flux in U band for the LRG spectrum).}
\label{fig:pass_spec_comp}
\end{figure*}

All the above composite spectra, normalized in the wavelength range 3950\AA{}\,$< \lambda <$\,4050\AA{} and smoothed with a three pixel boxcar filter, are plotted in the two panels of Fig.\,\ref{fig:pass_spec_comp}.

From this figure it appears that differences as well as similarities are present in our composite spectrum of passive galaxies residing in high-z dense environments and the other empirical templates considered here. In particular, we observe the following:
\begin{enumerate}
 \item The passive galaxies in the densest environments at $z \sim$ 1 exhibit a mean rest-frame U-V color that is redder than that of the coeval passive galaxies in K20, but bluer than that in groups at $z \sim$ 0. Assuming that passive galaxies of the two distant samples span the same metallicity range, those in clusters seem to have evolved faster than their counterparts in the field. However, because passive galaxies in the XDCP are more luminous (massive) than those in K20, the difference in their mean rest-frame U-V colors could be driven entirely by mass and not (also) by the density environment \citep[cf.][]{Thomas2005a, Thomas2010}. A more detailed spectroscopic analysis of our entire galaxy sample, aimed to characterize its star formation history, is in progress, however, and will be extensively discussed in a forthcoming paper.
\item The spectroscopic similarities between LRG and our passive galaxies would analogously suggest for the former sample an evolutionary path strongly driven by their \textit{mass} and likely characterized by different assembly histories.
\end{enumerate}
In summary, we find significant differences of our passive composite spectrum compared to similar ones in dense low-z environments as well as high-z field environments. The closest resemblance of our passive spectrum is found to the LRG template, obtained at intermediate redshift for very massive elliptical and bulge-dominated galaxies. All these findings seem to point toward a scenario where the evolutionary path of the passive galaxies, also residing in clusters, is mainly driven by their mass rather than environment.

\begin{figure}[t]
\centering
\vspace{-0.3cm}
\includegraphics[height=9.5cm, clip=true]{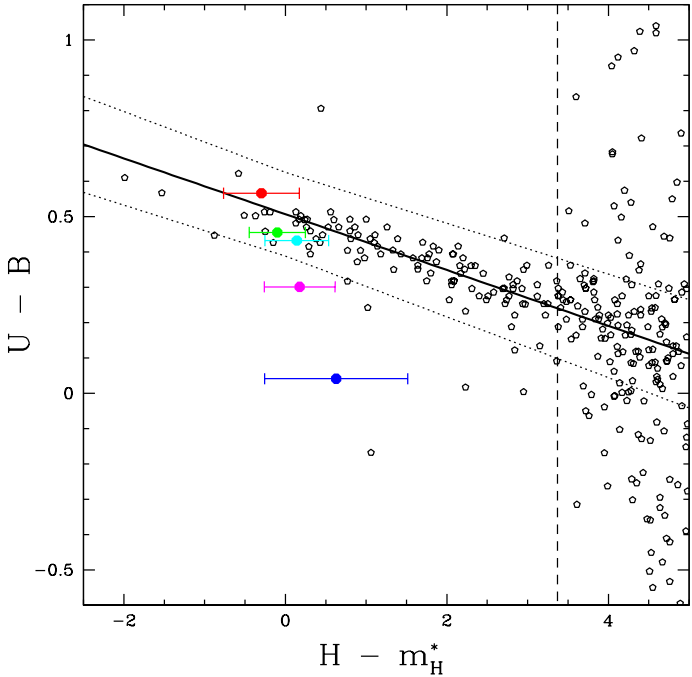} 
\caption{CMR of Coma Cluster galaxies (black pentagons) and the red sequence best fit, computed only for the early-type galaxies defined as brighter than H = 14.5 (at the left of the dashed vertical line), with relative scatter (black dotted lines) as provided by \cite{Eisenhardt2007}. The overplotted colored points mark the values obtained for the average spectra representative of the high-z cluster members, with the same color code as defined in Fig.\,\ref{fig:OII_Hdelta}. Note that for a consistent comparison all the observed H-band magnitudes were rescaled to the corresponding m$_H^{\star}$ values, as explained in the text. For Coma galaxies a value of m$_{H,Coma}^{\star}$ = 11.13 was used.\vspace{-0.3cm}}
\label{fig:U_B_Coma}
\end{figure}
\subsection{$U-B$ color of the templates}
\label{par:U_B_Coma_comparison}
In this section we compute the ($U-B$) rest-frame colors of the average spectra shown in Fig.\,\ref{fig:stackSpectra} and compare them to those typical of nearby galaxy clusters. Specifically, we refer to the values observed for the Coma Cluster ($z =$ 0.024) that were reported in the ($U-B$) vs H color-magnitude relation (CMR) of Fig.3 in \cite{Eisenhardt2007}.

Firstly, we obtained the ($U-B$) rest-frame color [($U-B$)$_{rest}$] for each spectroscopic class by convolving the corresponding rest-frame spectrum with the response curves of the U and B filters of the KPNO 0.9m telescope. We also computed the quoted \textit{observed} color by redshifting the average spectra to $z$ = 0.024 [($U-B$)$_{z=0.02}$] to obtain quantities comparable with Coma. These values are reported in the second and third column of Table\,\ref{Tab:U_B_values}, together with the differences between each rest-frame color and that typical of the passive class, taken as reference [$\Delta$($U-B$)$_{rest}$\textbar$_{passive}$, fourth column].

\begin{table*}[t]  
\centering          
\caption{($U-B$) colors of the five spectral templates at rest-frame and $z$ = 0.024 (second and third column, respectively). For each class the differences between its ($U-B$)$_{rest}$ color and the corresponding value of a passive galaxy is reported in the fourth column.}
\begin{tabular}{l c c c}
\vspace{-0.3cm} \\
\hline\hline       
\vspace{-0.3cm}            \\
Spectroscopic Class 	& ($U-B$)$_{rest}$& ($U-B$)$_{z=0.02}$ & $\Delta$($U-B$)$_{rest}$\textbar$_{passive}$\\ 
\vspace{-0.3cm}            \\
\hline    
\vspace{-0.3cm}            \\               
Passive			&  0.553 	 & 0.566	&  0.000  \\  
Post-starburst 		&  0.409 	 & 0.455	& -0.144  \\
Quiescent star-forming 	&  0.413 	 & 0.432	& -0.140  \\
Dusty starburst		&  0.229 	 & 0.301	& -0.324  \\
Starburst		& -0.027 	 & 0.041	& -0.580  \\
\vspace{-0.3cm}            \\
\hline \hline    \\              
\end{tabular}
\label{Tab:U_B_values} 

\caption{The median apparent Vega H-magnitude (Col. 4) together with the number of used galaxies and their median redshift (Cols. 2 and 3, respectively) for each spectral class. In Col. 5 the median of the difference between $H_{obs}$ and the apparent H magnitude of a L$^{\star}$ galaxy (at $z$\,=\,0) passively evolving since $z_{form}$ = 5 is reported. The errors refer to the semi-interquartile range.}
\begin{tabular}{l c c c c}
\vspace{-0.3cm} \\
\hline\hline     
\vspace{-0.3cm}            \\
Spectroscopic Class 	& N$_{gal}$		     & $<z>$ & $<H_{obs}>$ & $<H_{obs} - m_H^{\star}>$\\ 
\vspace{-0.3cm}            \\
\hline                    
\vspace{-0.3cm}            \\
passive			&  80	& 0.94 $\pm$ 0.07	& 18.49 $\pm$ 0.46 & -0.30 $\pm$ 0.47\\  
post-starburst 		&  11	& 1.07 $\pm$ 0.12	& 19.19 $\pm$ 0.26 & -0.10 $\pm$ 0.35\\
quiescent star-forming 	&  14	& 0.96 $\pm$ 0.02	& 19.52 $\pm$ 0.39 &  0.18 $\pm$ 0.44\\
dusty starburst		&  18	& 0.83 $\pm$ 0.11	& 18.71 $\pm$ 0.61 &  0.14 $\pm$ 0.39\\
starburst		&  10	& 0.97 $\pm$ 0.14	& 19.48 $\pm$ 0.55 &  0.63 $\pm$ 0.89\\
\vspace{-0.3cm}            \\
\hline \hline 
\end{tabular}
\label{Tab:H_mag} 
\end{table*}
After computing the five ($U-B$)$_{z=0.02}$ colors, we associated a median observed Vega H-magnitude ($<H_{obs}>$) to each spectroscopic class to then plot these values in a color-magnitude diagram. We defined $<H_{obs}>$ for each class as the median value of the apparent H-mag of the galaxies belonging to the class itself. However, only 13 out of the 16 XDCP clusters listed in Table\,\ref{Tab:clusterList} were observed in H-band, providing a total of 133 galaxies usable to this aim. The number of the galaxies with H-band data for each class, their median redshift and $H_{obs}$ are provided in Table\,\ref{Tab:H_mag}. 
Finally, to compare the computed values with those of Coma galaxies in a consistent way, we rescaled all observed H magnitudes to a characteristic m$_H^{\star}(z)$. It is defined as the apparent H magnitude of a galaxy at redshift $z$ that has L = L$^{\star}$ at $z$ = 0, passively evolving since $z_{form} = $ 5. We adopted a suite of spectral energy distribution models\footnote{In particular, these models assume a \cite{Salpeter1955} stellar initial mass function with lower and upper mass cut-offs equal to 0.1 and 120 M$_{\odot}$, fixed solar metallicity and a range of 130 epochs with a time step of 0.1 Gyr since the instantaneous burst of formation. This choice provides a suitable representation of the optical/near-infrared colors of observed, red, and dead galaxies associated with old, passively evolving stellar populations and an elliptical morphology in different density environments up to $z \sim 2$ \citep[e.g., ][]{Pierini2004, Wilman2008, Fassbender2011b}.}, corresponding to the evolution of a simple stellar population, computed within the framework of PEGASE 2 \citep{Fioc1997}. The stellar mass-scale is matched to the observed magnitude of an $L^{\star}$ galaxy in clusters at $z \sim 0$ in the Ks-band \citep{Strazzullo2006a}. The association between age of the model and redshift of the observed galaxy is set by the adopted cosmological model. 

Since m$_H^{\star}$ is a function of redshift, the appropriate value was computed and subtracted from $H_{obs}$ for each galaxy according to its redshift. The median $<H_{obs} - m_H^{\star}>$ was finally obtained for each spectral class and is quoted in the fourth column of Table\,\ref{Tab:H_mag}.
The apparent H magnitudes of Coma galaxies were equivalently rescaled by referring to m$_{H,Coma}^{\star}$ = 11.13 as quoted by \cite{Eisenhardt2007} and found by \cite{DePropris1998}.

The final ($U-B$)$_{z=0.02}$ vs  $H_{obs} - m_H^{\star}$ CMR for Coma and the average spectra of distant cluster members is shown in Fig.\,\ref{fig:U_B_Coma}.

From Fig.\,\ref{fig:U_B_Coma} it is clear how the ($U-B$) colors of the Coma early-type galaxies are fully consistent with the typical ($U-B$) colors of the 0.6 $\leq z \leq$ 1.2 cluster members classified as passive, post-starburst, and quiescent star-forming from our analysis. This result may indicate that the stellar mass formation of the local passive cluster galaxies was almost completed already at $z >$ 0.8, so that they did not experience any additional subsequent significant star formation process that could have made them bluer at $z \sim$ 0. This appears to be confirmed because the values of EW(H$\delta _A$) and D$_n$4000 (reported in Table\,\ref{Tab:specIndex_Templates}) of the composite spectrum of distant passive galaxies are similar to those found for the passive and most massive galaxies in the local universe. These findings suggest that these galaxies have created the bulk of their stellar mass already at $z >$ 0.8 and, according to the results discussed in Sect.\,\ref{par:passive_spec_comparison}, possibly even faster than the coeval galaxies in the field. As mentioned in Sect.\,\ref{par:passive_spec_comparison}, another study on the age-metallicity relation of the passive galaxies contained in our sample is currently underway and will be extensively discussed in a forthcoming paper.

Finally, Fig.\,\ref{fig:U_B_Coma} also shows that the starburst galaxies at $z \leq$ 0.6 appear to be more luminous (in H-band) than the Coma galaxies with the same color. Even though a large part of this effect may be due to the expected selection bias toward the brightest objects in the distant universe, that in Coma very few star-forming objects appear to have H $\sim$ m$_{H,Coma}^{\star}$ is a consequence of the \textit{downsizing} scenario according to which the most vigorous star-formation activity at high redshift was actually located in the most massive galaxies \citep[e.g.,][]{Cowie1996a, Gavazzi1996a, Gavazzi1996b, Juneau2005, Pannella2009}. This effect is also believed to have played a fundamental role in building-up the red sequence in galaxy clusters starting from the more massive galaxies \citep[see e.g.,][]{Romeo2008}.

\section{Conclusion}
\label{par:discussion}
We presented \textit{F}-VIPGI, a new adapted version of VIPGI for reducing FORS2 spectroscopic data in a semi-automated and efficient way. We discussed the major improvements and technical aspects of the new software, which is now available to the scientific community, in the first part of the paper. Additional information on the presented software are provided in Appendix\,A.

In the second part we showed an application of the above pipeline to a sample of distant (0.65 $\leq z \leq$ 1.25) X-ray selected galaxy clusters as part of the XDCP sample. We classified all galaxies according to their spectral indices EW([OII]) and EW(H$\delta$) and stacked their spectra to create a new library of templates, which is particularly suited for a better redshift (spectroscopic and photometric) measurements of galaxies residing in dense high-z environments.

In the last part of the paper we compared the composite passive spectrum with others representative of the passive galaxy populations residing in different environments and cosmic epochs, finding some remarkable differences and similarities among them. 

Namely, we found that passive galaxies in clusters appear to be more evolved already at $z \sim$ 0.8 w.r.t. the field galaxies at similar redshift, supporting the idea that galaxy evolution is significantly accelerated in the densest environments. However, because our sample is significantly biased toward higher masses, we cannot exclude the importance of the \textit{mass} as another driving element of evolution either. We finally studied the ($U-B$) colors of the entire sample of composite spectra and compared them to those observed in the Coma Cluster. The results of this work show that the colors of the passive galaxies in local clusters are fully consistent with those observed for distant cluster members classified as passive, post-starburst, and quiescent starforming according to our criteria. This suggests that the stellar mass formation in local cluster galaxies was almost completed already at $z \sim$ 0.8. In addition, comparing the mean H-band luminosities of the bluest galaxies of our sample with those in Coma, we found that at $z \geq$ 0.65 the star-forming galaxies are much more luminous. Beyond the expected selection bias, the observed effect may be partly explained by assuming a \textit{downsizing} scenario, where the most vigorous starforming galaxies in the far universe were also the most massive ones.

\appendix
\section{Additional information on \emph{F}-VIPGI}
\label{par:Appendix_A}
In the following appendix we provide some additional information useful to readers who consider using the software discussed in this paper.

\subsection{Compatibility with the current operating systems}
\label{par:Appendix_A_compatibility}
\textit{F}-VIPGI is fully compatible with those architecture softwares where the old version of VIPGI can be installed and run. Since a few years, however, the plotting library Pmw.BLT is not updated anymore and hence started to become obsolete and incompatible with some recent operating systems and with all current 64-bit machines. 

In the following we list all operating systems that fully support the current version of \textit{F}-VIPGI with their version numbers reported in parenthesis:
\begin{itemize}
\item[$\bullet$] Linux SuSE (9.X, 10.X, 11.0 and 11.1)
\item[$\bullet$] Linux Fedora (from 8 to 14)
\item[$\bullet$] Linux Ubuntu (from 8.xx to 10.xx)
\item[$\bullet$] Solaris (2.5)
\item[$\bullet$] Mac OSX (10.5 and 10.6)
\end{itemize}

Another improved version of VIPGI is already being developed and is expected to be completed by the mid of 2013.
It will use a completely renewed graphical interface, fully usable also with the most recent architecture softwares, and it will also include reduction recipes suitable for LUCIFER data, the spectrograph mounted at the Large Binocular Telescope.

We finally stress that \textit{F}-VIPGI is publicly released to the community as it is and without any guarantee of technical support due to the lack of funds to support such assistance.

\subsection{How to obtain \textit{F}-VIPGI}
\label{par:Appendix_A_getFVIPGI}
The add-ons and the binary files of \textit{F}-VIPGI can be downloaded from the following link:
\begin{center}
\href{http://www.mpe.mpg.de/cosmology/cluster/FVIPGI}{http://www.mpe.mpg.de/cosmology/cluster/FVIPGI}
\end{center}

In the package, the cookbook of the new pipeline is also included.

\subsection{Link for spectroscopic templates}
\label{par:Appendix_A_getLibrary}
The new library of spectra described in this manuscript, as well as some related technical information, are available in electronic form as ASCII and Flexible Image Transport System (\textit{fits}) files at the CDS and also via
\begin{center}
\href{http://www.mpe.mpg.de/cosmology/cluster/FVIPGI/specLib}{http://www.mpe.mpg.de/cosmology/cluster/FVIPGI/specLib}
\end{center}
\newpage
\begin{acknowledgements}
We thank the anonymous referee for the insightful comments that helped to improve the clarity of the paper.
The \XMM project is an ESA Science Mission with instruments and contributions directly funded by ESA Member States and the USA (NASA).
The XMM-Newton project is supported by the Bundesministerium f\"ur Wirtschaft und Technologie/Deutsches Zentrum f\"ur Luft- und Raumfahrt (BMWI/DLR, FKZ 50 OX 0001), the Max-Planck Society and the Heidenhain-Stiftung. 
This research has made use of the NASA/IPAC Extragalactic Database (NED) which is operated by the Jet Propulsion Laboratory, California Institute of Technology, under contract with the National Aeronautics and Space Administration. 
This work was supported by the Munich Excellence Cluster Origin and Structure of the Universe (www.universe-cluster.de), by the DFG under grants Schw536/24-1, Schw 536/24-2, BO 702/16-1,16-2,16-3, the Transregio TR33 and the German DLR under grant 50 QR 0802. AN is grateful to Angela Bongiorno and Italo Balestra for fruitful discussions. DP acknowledges the kind hospitality at the MPE.
\end{acknowledgements}

\bibliographystyle{aa}
\bibliography{aa19862_biblio}
\end{document}